\documentclass[preprint2]{proto}
\usepackage{times}
\newcommand{\refs}{\par\noindent\hangindent=1pc\hangafter=1}
\begin{document}

\title{\textbf{\LARGE 
Evolution of Circumstellar Disks Around Normal Stars: \break 
Placing Our Solar System in Context}}

\author {\textbf{\large Michael R. Meyer}}
\affil{\small\em The University of Arizona}
\author {\textbf{\large Dana E. Backman}}
\affil{\small\em SOFIA/SETI Institute}
\author {\textbf{\large Alycia J. Weinberger}}
\affil{\small\em Carnegie Institution of Washington}
\author {\textbf{\large Mark C. Wyatt}}
\affil{\small\em University of Cambridge}

\begin{abstract}
\baselineskip = 11pt
\leftskip = 0.65in 
\rightskip = 0.65in
\parindent=1pc
Over the past 10 years abundant evidence has emerged that many (if not 
all) stars are born with circumstellar disks.   Understanding the 
evolution of post--accretion disks can provide strong 
constraints on theories of planet formation and evolution.  
In this review, we focus on developments in understanding:
a) the evolution of the gas and dust content of circumstellar disks 
based on observational surveys, highlighting new results from the 
Spitzer Space Telescope; b) the physical properties
of specific systems as a means to interpret the survey 
results; c) theoretical models used to explain the observations; d) 
an evolutionary model of our own solar system for comparison to the
observations of debris disks around other stars; and e) how these new results
impact our assessment of whether systems like our own are common or rare 
compared to the ensemble of normal stars in the disk of the Milky Way.
\\~\\~\\~%}%leave this in to get the correct vertical space after the abstract

\end{abstract}

\section{Introduction}

At the first Protostars and Planets conference in 1978, the
existence of circumstellar disks around sun--like stars was
in doubt, with most researchers preferring the hypothesis 
that young stellar objects were surrounded by spherical 
shells of material unlike the solar nebula thought to give
rise to the solar system ({\it Rydgren et al.}, 1978).  
By the time of Protostars and 
Planets II, experts in the field had accepted
that young stars were surrounded by circumstellar
disks though the evidence was largely circumstantial 
({\it Harvey}, 1985).  At that meeting, Fred Gillett and 
members of the IRAS team announced details of newly 
discovered debris disks, 
initially observed as part of the calibration program 
({\it Aumann et al.}, 1984).   At PPIII, it was well--established
that many stars are born with circumstellar accretion disks
({\it Strom et al.}, 1993) and at PPIV, it was recognized that many of these disks 
must give rise to planetary systems ({\it Marcy et al.}, 2000). 
Over the last 15 years, 
debris disks have been recognized as playing an important 
role in helping us understand the formation and evolution
of planetary systems 
({\it Backman and Paresce}, 1993; {\it Lagrange et al.}, 2000; 
see also {\it Zuckerman}, 2001).
After PPIV, several questions remained.  How do debris
disks evolve around sun--like stars?  When do gas--rich 
disks transition to debris disks?  Can we infer the 
presence of extra--solar planets from spectral energy
distributions (SEDs) and/or resolved disk morphology?  Is there 
any connection between debris disks and the radial velocity
planets?  Is there evidence for differences in disk 
evolution as a function of stellar mass? 

In answering these questions, 
our objective is no less than to understand the formation and evolution of 
planetary systems through observations of the gas and dust
content of circumstellar material surrounding stars
as a function of stellar age.  By observing how disks dissipate
from the post--accretion phase through the planet building
phase we can hope to constrain theories of planet formation (cf. 
chapters by {\it Durisen et al.} and 
{\it Lissauer and Stevenson}).  By 
observing how debris disks generate dust at late times
and comparing those observations with physical models
of planetary system dynamics, 
we can infer the diversity of solar system
architectures as well as attempt to understand 
how they evolve with time. 

Today, we marvel at the wealth of results from the Spitzer Space
Telescope and high contrast images of spectacular individual 
systems. Detection statistics that were very uncertain with 
IRAS and ISO sensitivity now 
can be compared with models of planetary system 
evolution, placing our solar system in context.  Advances in 
planetary system dynamical theory, the discovery and characterization of 
the Kuiper Belt (see chapter by {\it Chiang et al.})
have proceeded in parallel and further contribute to our understanding 
of extrasolar planetary systems.  We attempt to compare 
observations of disks surrounding other stars to our current
understanding of solar system evolution.  
Our ultimate goal is to learn whether or not solar systems
like our own are common or rare among stars in the 
disk of the Milky Way and what implications this might have
on the frequency of terrestrial planets that might give 
rise to life. 

Our plan for this contribution is as follows. 
In Section 2, we describe recent results from observational 
surveys for gas and dust surrounding normal stars.  Next
we describe detailed studies of individual objects in Section 3.
In Section 4, we review modeling approaches used in constraining
physical properties of disks from the observations.  Section 5
describes a toy model for the evolution of our solar system
which we use to compare to the ensemble of observations.  Finally, 
in Section 6 we attempt to address whether or not planetary systems
like our own are common or rare in the Milky Way galaxy 
and summarize our conclusions. 

\section{Evolution of Circumstellar Disks} 

In order to study the evolution of circumstellar
disks astronomers are forced to observe sun--like
stars at a variety of ages, in an attempt to create
a history, hoping that on average, a younger population
of similar mass stars can be assumed to be the evolutionary
precursors of the older.  Although deriving ages of stars
across the H--R diagram is fraught with uncertainty 
(e.g., {\it Stauffer et al.}, 2004) it is a necessary
step in studies of disk evolution.  Such studies, combined
with knowledge of our own solar system, are the only observational 
tools at our disposal for constraining theories of planet formation. 

\subsection{Statistics from Dust Surveys} 

{\em Circumstellar Dust within 10 AU} 
Nearly all stars are thought to be born with circumstellar
disks ({\it Beckwith and Sargent}, 1996; 
{\it Hillenbrand et al.}, 1998)
and recent work has shown that these disks dissipate on timescales 
of order 3 Myr ({\it Haisch et al.}, 2001).  However, these
results are based largely on the presence of near--infrared
excess emission which only traces optically--thick hot dust 
within 0.1 AU of the central star.  Indeed the
presence of an inner disk appears to correlate with the presence or 
absence of spectroscopic signatures of active accretion onto
the star ({\it Hartigan et al.}, 1995; see also chapter by 
{\it Bouvier et al.}).  As active
disk accretion diminishes ({\it Hartmann et al.}, 1998), 
the fraction of young stars in clusters that show 
evidence for optically--thick inner disks diminishes. 
Yet what is often overlooked is
that the very data that suggest a typical 
inner disk lifetime of $\sim$ 3 Myr, also {\it suggests
a dispersion of inner disk lifetimes from 1--10 Myr}.

What has remained unclear until recently is how these primordial disks 
left over from the formation of the young star dissipate at
larger radii and whether the termination of accretion represents
an end of the gas--rich phase of the circumstellar disk. 
Even at the time of PPIII, it was recognized that
young stars (with ages $<$ 3 Myr) 
lacking optically--thick near--infrared excess
emission but possessing optically--thick mid--infrared emission
were rare ({\it Skrutskie et al.}, 1990).  This suggested that 
the transition time between optically--thick and thin from 
$<$ 0.1 AU to $>$ 3 AU was rapid, $<<$ 1 Myr 
({\it Wolk and Walter}, 1996; 
{\it Kenyon and Hartmann}, 1995; {\it Simon and Prato}, 1995).  

It is important to distinguish between surveys for primordial 
disks, gas and dust rich disks left over from the star formation
process, and debris disks, where the opacity we see is 
dominated by grains released through collisions of larger
parent bodies.  Often this distinction is made based on 
whether remnant gas is left in the system.  With a gas to 
dust ratio $>$ 1.0, dust dynamics are influenced by 
their interaction with the gas ({\it Takeuchi and Artymowicz}, 2001). 
In the absence of gas, one can argue based on the short dust
lifetimes that observed dust is likely recently generated
through collisions in a planetesimal belt ({\it Backman and Paresce}, 1993; 
{\it Jura et al.}, 1998).  Observations that constrain evolution of the 
gas content in disks are described below.

Recent
work has shown that even optically--thin mid--infrared 
emission (tracing material between 0.3--3 AU) is rare
around sun--like stars with ages 10--30 Myr.  {\it Mamajek 
et al.} (2004) performed a survey for excess emission 
around sun--like stars in the 30 Myr old Tucana--Horologium
association and found no evidence for excess within a 
sample of 20 stars down to dust levels $<$ 2 $\times 10^{-6}$
M$_{\oplus}$ for warm dust in micron--sized grains.  Similar
studies by {\it Weinberger et al.} (2004) of stars in the $\beta$
Pic moving group as well as TW Hya association (both $\sim$ 10 Myr
old) uncovered only a handful of stars with mid--infrared excess
emission.  These results are being confirmed with cluster studies 
undertaken with the Spitzer Space telescope.  As 
part of the {\it Formation and Evolution of Planetary 
Systems} (FEPS) Legacy Science Program 
a survey has been conducted searching for warm dust
at wavelengths from 3.6--8.0 $\mu$m around 74 
sun--like
stars with ages 3--30 Myr.  {Silverstone et al.} (2006) 
reported only five detections from this survey and all of those 
were rare examples of long--lived optically--thick disks. 
{\it It appears that circumstellar disk material 
between 0.1--1 AU typically drops below detectable levels on 
timescales comparable to the cessation of accretion.} 
These levels are probably below what our solar system might have 
looked like at comparable ages (3--30 Myr). 

However, Spitzer is uncovering a new population of transitional 
disks at mid--infrared wavelengths 
in the course of several young cluster surveys 
({\it Forrest et al.}, 2004; {\it Calvet et al.}, 2005).  
{\it Chen et al.} (2005) 
find that $\sim$ 30 \% of sun--like stars in the subgroups of the 
5--20 Myr Sco Cen OB association exhibit 24 $\mu$m 
excess emission, higher than that found by {\it Silverstone 
et al.} (2006) at shorter wavelengths.  {\it Low et al.} 
(2005) find examples of mid--IR excess at 24 $\mu$m 
in the 10 Myr TW Hya association.  
The 24 $\mu$m emission is thought to trace material $>$ 1 AU, 
larger radii than the material traced by emission from 
3--10 $\mu$m.  Preliminary results from the FEPS
program suggests that there is some evolution 
in the fraction of sun--like stars with 24 $\mu$m excess 
(but no excess in the IRAC bands) from 3--300 Myr. 
This brackets major
events in our own solar system evolution with the terrestrial 
planets thought to have mostly formed in $<$30 Myr and the
late heavy bombardment at $>$300 Myr (see Section 5). 

It is interesting to note that there is now a small (5 member) class
of debris disks with only strong mid--infrared excess and weak 
or absent far--IR/sub--mm excess emission: BD +20$^o$307 at
$>$300 Myr age ({\it Song et al.}, 2005), HD 69830 at $\sim$2 Gyr age
({\it Beichman et al.}, 2005b), HD 12039 at 30 Myr 
({\it Hines et al.}, 2006), HD 113766 
at 15 Myr ({\it Chen et al.}, 2005), and HD 98800 at 10 Myr 
({\it Low et al.}, 1999; {\it Koerner et al.}, 2000). 
In the two older systems, BD +20$^o$307 and 
HD 69830, this excess is almost entirely
silicate emission from small grains.  These objects are rare, 
only 1--3 \% of all systems surveyed.  Whether they represent 
a short--lived transient phase that all 
stars go through, or a rare class of massive warm 
debris disks is not yet clear (Section 4.4). 

{\em Circumstellar Disks at Radii $>$ 10 AU} Surveys at far--infrared ($>$ 30 $\mu$m) and sub--millimeter
wavelengths trace the coolest dust at large radii.  Often, 
this emission is optically--thin and is therefore a good
tracer of total dust mass at radii $>$ 10 AU.  Early surveys 
utilizing the IRAS satellite focused on large optically--thick 
disks and envelopes surrounding young stellar objects 
within 200 pc, the distance of most star--forming regions
({\it Strom et al.}, 1993), 
and main sequence stars within 15 parsecs 
because of limitations in sensitivity 
({\it Backman and Paresce}, 1993). Sub--millimeter
work suggested that massive circumstellar disks 
dissipate within 10 Myr ({\it Beckwith et al.}, 1990; 
{\it Andrews 
and Williams}, 2005).  Sub--millimeter surveys of field stars indicated that 
``typical'' sub--millimeter emission from dust surrounding main 
sequence stars diminished as $t^{-2}$ ({\it Zuckerman and Becklin}, 1993). 

Several new far--infrared studies were initiated with the launch of the 
Infrared Space Observatory (ISO) by ESA and the advent of 
the sub--millimeter detector SCUBA on the JCMT.  
{\it Meyer and Beckwith} (2000) describe surveys of young clusters 
with the ISOPHOT instrument on ISO which indicated that
far--infrared emission became optically--thin on timescales
comparable to the cessation of accretion (about 10 Myr). 
{\it Habing et al.} (1999, 2001) 
suggested that there was another discontinuity in the evolutionary
properties of debris disks surrounding isolated A stars at an age of 
approximately 400 Myr.  {\it Spangler et al.} (2001) conducted a large survey 
including both clusters and field stars finding that dust mass 
diminished at $t^{-1.8}$ as if the dust removal mechanism was
P--R drag (see Section 4 below).  Based on the data 
available at the time, and limitations in sensitivity from ISO, 
it was unclear how to reconcile these disparate conclusions
based on comparable datasets.  For a small sample of sun--like
stars, {\it Decin et al.} (2003)
found that 10--20 \% ($5/33$) of Milky Way G stars, 
regardless of their age, have debris disks, comparable to results
obtained previously for A stars ({\it Backman \& Paresce}, 1993). 

Recent work with the Spitzer Space Telescope offers a new perspective.
From the FEPS program, surveys for cold debris disks surrounding G stars
have led to several new discoveries 
({\it Meyer et al.}, 2004; {\it Kim et al.}, 
2005).   Over 40 debris disk candidates have been 
identified from a survey of 328 stars and no strong correlation of 
cold dust mass with stellar age has been found. 
{\it Bryden et al.} (2006; see also {\it Beichman et al.}, 2005a) have 
completed a volume--limited survey of nearby sun--like stars
with probable ages between 1--3 Gyr old.  Overall the Spitzer statistics 
suggest a cold debris disk frequency of 10--20 \% surrounding
sun--like stars with a weak dependence on stellar age 
(Fig. 1).  It should
be noted that our own solar system cold dust mass would be 
undetectable in these surveys and it is still difficult to assess
the mean and dispersion in cold disk properties based on the 
distribution of upper limits. 

Sub--millimeter surveys of dust mass probe the coldest 
dust presumably at the larger radii.  {\it Wyatt et al.}
(2003) report observations of low mass companions to 
young early--type stars (see also 
{\it Jewitt et al.}, 1994) indicating a lifetime of 10--60 Myr for
the massive primordial disk phase.  {\it Carpenter et al.} (2005; 
see also {\it Liu et al.}, 2004), 
combined these data with a new survey from the FEPS sample
and found that the distribution of dust masses (and upper limits)
from 1--3 Myrs is distinguished (with higher masses) 
than that found in the 10--30 Myr 
old sample at the 5 $\sigma$ level (Fig 1).  The data do not permit 
such a strong statement concerning the intermediate age 
3--10 Myr sample.  
{\it Najita and Williams} (2005) conducted a detailed study of
$\sim$ 15 individual objects and find that debris disks do not
become colder (indicating larger radii for the debris)
as they get older surrounding sun--like stars in contrast
to the predictions of {\it Kenyon and Bromley} (2004).  Again 
we note that these surveys would not detect the sub-mm
emission from our own Kuiper Debris Belt (see Section 5 
below).  In contrast, {\it Greaves et al.} (2004) point out that
the familiar tau Ceti is 30 times more massive than our solar
system debris disk, even at comparable ages. 
{\it Greaves et al.} (2006) studied the 
metallicities of debris disk host stars showing that their distribution 
is indistinguishable from that of field stars in contrast to the exoplanet 
host stars which are metal-rich ({\it Fischer and Valenti}, 2005). 
Implications of the detected debris disk dust masses 
and their expected evolution is discussed in Section 4
and compared to the evolution of our solar system in Section 5. 

The picture that emerges is complex as illustrated in Fig. 1. 
In general, we observe diminished cold dust mass with time
as expected from models of the collisional evolution of
debris belts (see Section 4).  However, at any one age there
is a wide dispersion of disk masses.  Whether this dispersion
represents a range of initial conditions in disk mass, a range
of possible evolutionary paths, or is evidence that many disks
pass through short--lived phases of enhanced dust production
is unclear.  One model for the evolution of our solar system
suggests a rapid decrease in observed dust mass associated
with the dynamic rearrangement of the solar system at 700 Myr
(and decrease in the mass of colliding parent bodies by $\times$ 10). 
If that model is correct, we would infer that our solar
system was an uncommonly bright debris disk at early times, 
and uncommonly faint at late times (see Section 6). 

\begin{figure*}
\epsscale{1.0}
\plotone{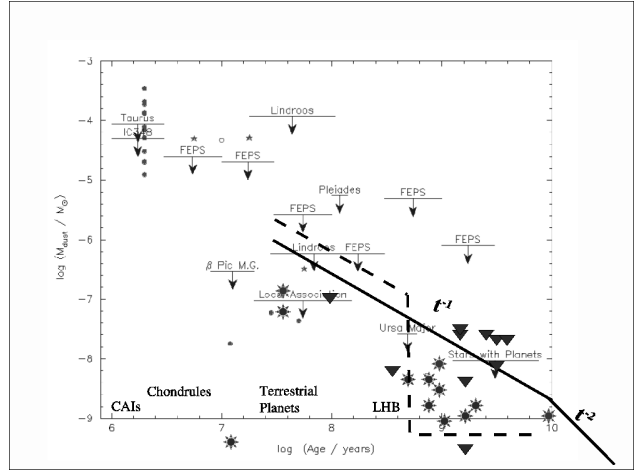}
\caption{\small Evolution of circumstellar dust mass 
based on sub--mm observations from {\it Carpenter et al.} (2005). 
Over--plotted are 
Spitzer 70 $\mu$m detections from the FEPS program 
(stars) and upper limits (triangles). 
Slopes of t$^{-1}$ and t$^{-2}$ are shown as solid lines, 
along with a toy model for the evolution of our solar system 
(denoted with a dashed line) indicating an abrupt transition in dust mass associated
with the late--heavy bombardment (LHB). Timescales associated
with the formation of calcium--aluminum inclusions (CAIs), chondrules, 
and terrestrial planets are also shown.}
 \end{figure*}

\subsection{Statistics from Gas Surveys} 

While most energy is focused on interpreting dust
observations in disks, it is the gas that dominates
the mass of primordial disks and is the material
responsible for the formation of giant planets. 
Observational evidence for the dissipation of gas
in primordial disks surrounding young sun--like stars
is scant.  Millimeter wave surveys (see chapter by {\it Dutrey et al.}) 
are on--going and confirm the basic results:
1) classical T Tauri stars with excess emission 
from the near--IR through the sub--millimeter are gas
rich disks with some evidence for Keplerian support; 
and 2) complex chemistry and gas--grain interactions
affect the observed molecular abundances.  In a pioneering
paper, {\it Zuckerman et al.} (1995) suggested that gas rich 
disks dissipate within 10 Myr.  Recent work on disk 
accretion rates of material falling ballistically
from the inner disk onto the star 
by {\it Lawson et al.} (2004) could
be interpreted as indicating gas--rich primordial 
disks typically dissipate on timescales of 3--10 Myr. 
Other approaches include observations of warm 
molecular gas through near--infrared spectroscopy 
(see chapter by {\it Najita et al.}), UV absorption line
spectroscopy of cold gas for favorably 
oriented objects (see next section), and mm--wave
surveys for cold gas in remnant disks. 
One debris disk that showed evidence for gas in the early
work of {\it Zuckerman et al.} (1995), the A star 49 Ceti, was recently
confirmed to have CO emission by {\it Dent et al.} (2005). 
Transient absorption lines of atomic gas with 
abundances enhanced in refractory species would 
suggest the recent accretion of comet--like material
({\it Lecavelier des Etangs et al.}, 2001). 

Since most of the mass in molecular clouds, and presumably
in circumstellar disks from which giant planets form is
molecular hydrogen, it would be particularly valuable
to constrain the mass in H$_2$ directly from observations. 
ISO provided tantalizing detections of warm H$_2$ at 
12.3, 17.0, and 28.2 $\mu$m tracing gas from 50--200 K
in both primordial and debris disks ({\it Thi et al.}, 2001a,b).  
However follow--up observations with high 
resolution spectroscopic observations (with a much 
smaller beam--size) have failed to 
confirm some of these observations ({\it Richter et al.}, 2002; 
{\it Sheret et al.}, 2003).  Several surveys for warm molecular gas are underway
with the Spitzer Space Telescope.  {\it Gorti and Hollenbach}
(2004) present a series of models for gas rich disks
with various gas to dust ratios.  The initial stages
of grain growth in planet forming disks, the subsequent
dissipation of the primordial gas disk, and the onset of 
dust production in a debris disk suggest a wide range
of observable gas to dust ratios (see the chapter by 
{\it Dullemond et al.}).  {\it Hollenbach et al.} 
(2005) placed upper limits of 0.1 M$_{JUP}$ to the gas content 
of the debris disk associated with HD 105, a 30 Myr old
sun--like star observed as part of the FEPS project. 
{\it Pascucci et al.} (submitted) have presented results for 
a survey finding no gas surrounding 15 stars with ages
from 5--400 Myr (nine of which are younger than 30 Myr) 
at levels comparable to HD 105.  Either
these systems have already formed extra--solar giant
planets, or they never will.  Future work will concentrate
on a larger sample of younger systems with ages 1--10 Myr
in order to place stronger constraints on the timescale
available to form gas giant planets. 

\section{Physical Properties of Individual Systems}

In order to interpret results from the surveys described above, 
we need to understand in detail the composition and structure of
debris disks. Presumably, the dust (see Section 4.1) reflects the
composition of the parent planetesimal populations, so measuring 
the elemental composition, organic fraction, ice fraction, and
ratio of amorphous to crystalline silicates provides information on the
thermal and coagulation history of the small bodies. These small bodies
are not only the building blocks of any larger planets, they could be an 
important reservoir for delivering volatiles 
to terrestrial planets (e.g., {\it Raymond et al.}, 2004).
Additionally, the grain size distribution reflects the collisional state
of the disk (see Section 4.1).  The structure of the disk may reflect
the current distribution of planetesimals and therefore the system's
planetary architecture (see Section 5).

The literature on resolved images of circumstellar disks begins with
the pioneering observations of $\beta$ Pic by {\it Smith and Terrile} 
(1984).  Since PPIV, there has been a significant increase in spatially resolved
information on debris disks in two regimes -- scattering and emission.
Resolving scattered visual to near-infrared light requires high contrast
imaging such as that delivered by HST, because the amount of scattered
light is at most 0.5\% of the light from the star. Resolving thermal
emission requires a large aperture telescope because dust is warm closer
to the star and so disks appear quite small in the infrared.

Compositional information is obtained from scattered light albedos and
colors, from mid-infrared spectroscopy that reveals solid-state
features, and from fitting the slopes observed in spectral energy
distributions. Resolved imaging breaks degeneracies in disk model fits and
can be used to investigate changes in composition with
location. Structural information is best at the highest spatial
resolution and includes observations of warps, rings, 
non-axisymmetric structures, and offset centers of symmetry. 

Sensitivity to grain size depends on wavelength and each regime provides
information on grains within approximately a range of 0.1-10 times the
wavelength (Fig. 2). For example, scattered visible and near-infrared
light mostly probes grains smaller than 2 $\mu$m and submillimeter
emission mostly probes grains $>$ 100 $\mu$m in size.

\begin{figure*}
\epsscale{0.75}
\plotone{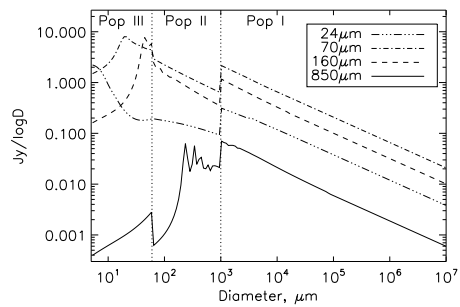}
\caption{\small Contribution of different grain sizes to the fluxes observed
in different wavebands in the Vega disk ({\it Wyatt}, 2006).  The units
of the y--axis are flux per log particle diameter so that the 
area under the curve indicates the contribution of different sized
particles to the total flux for a given wavelength.  The different wavebands
probe different ranges in the size distribution and so are predicted 
to see very different structures.
}
 \end{figure*}

\subsection{Debris Disks Resolved in Scattered Light}

The number of debris disks resolved in scattered light has increased
from one at the time of PPIII ($\beta$ Pic) to two at the time of PPIV
(HR 4796) to about 10 today (see Table 1). 
The detection of 55 Cnc reported in PPIV seems to be
spurious ({\it Schneider et al.}, 2001; {\it Jayawardhana et al.}, 2002), 
and HD 141569 is not gas-free ({\it Jonkheid et al.}, 2005) and 
therefore is not counted here as a debris disk. 

\begin{deluxetable}{lclllllll}
\tabletypesize{\small}
\tablecaption{Resolved Debris Disk Properties \label{tbl-1}}
\tablewidth{0pt}
\tablehead{
\multicolumn{9}{c}{Scattered Light}\\
Star         &Sp. &Age   &Size    &Color             &g    &albedo &Resolved &References\\
             &Typ.&(Myr) &(AU)    &                  &     &       &in Emis?  &
}
\startdata
HR 4796A     &A0  & 8    &70      &red (V-J)         &0.15 &0.1-0.3 &yes      &1,2,3,4\\
HD 32297     &A0  &10?   &400     &blue (R-J)        &Not Avail.  &0.5     &no&5,6\\
$\beta$ Pic  &A5  &12    &10-1000 &neutral-red (V-I) &0.3-0.5   &0.7&yes      &7,8,9,10\\
AU Mic	     &M1  &12    &12-200  &neutral-blue (V-H)&0.4  &0.3     &no	      &11,12,13,14\\
HD 181327    &F5  &12    &60-86   &Red (V-J)         &0.3  &0.5     &no	      &15 \\
HD 92945     &K1  &20-150&120-146 &Red (V-I)         &Not Avail.&Not Avail. &no &16 \\
HD 107146    &G2  &30-250&130     &red (V-I)         &0.3  &0.1     &yes      &17,18\\
Fomalhaut    &A3  &200   &140     &Not Avail.        &0.2  &0.05    &yes      &19,20\\
HD 139664    &F5  &300   &110  	  &Not Avail.	     &Not Avail.  &0.1     &no&21 \\
HD 53143     &K1  &1000  &110	  &Not Avail.	     &Not Avail.  &0.06    &no&21 \\
Saturn's Rings&--   &--  &--      &red (B-I)         &-0.3 &0.2-0.6 &--       &22 \\
\hline
\multicolumn{9}{c}{Emission (Additional)}\\
Vega          &A0  &200    &$>$90  &                  &     &        &	      &23,24,25\\
$\epsilon$ Eridani&K2 &$<$1000&60     &                  &     &       &      &26,27 \\
$\eta$ Corvi &F2  &$\sim$1000&100  &                  &     &        &	      &28\\
$\tau$ Ceti  &G8  &$\sim$5000&55  &                  &     &        &	      &29 \\
\enddata
\tablecomments{Notes: The size given is the approximate radius or range
of radii.  It remains to be seen if the younger systems, particularly HD 32297,
really are gas-free debris disks. The size for HD 32297 is the inner
disk; it has a large circumstellar nebulosity as well (Kalas 2005).}
\tablerefs{
1. {\it Schneider et al.} (1999), 
2. {\it Schneider, G. and Debes, J.} (personal communication), 
3. {\it Jayawardhana et al.} (1998),
4. {\it Koerner et al.} (1998),
5. {\it Schneider et al.} (2005),
6. {\it Kalas} (2005),
7. {\it Artymowicz et al.} (1989), 
8. {\it Kalas and Jewitt} (1995),
9. {\it Telesco et al.} (2005),
10. {\it Golimowski et al.} (2005),
11. {\it Kalas et al.} (2004),
12. {\it Liu} (2004), 
13. {\it Metchev et al.} (2005), 
14. {\it Krist et al.} (2005), 
15. {\it Schneider et al.} (in press),
16. {\it Clampin et al.} (in preparation),
17. {\it Ardila et al.} (2004),
18. {\it Williams et al.} (2004), 
19.  {\it Wyatt and Dent} (2002),
20.  {\it Kalas et al.} (2005),
21. {\it Kalas et al.} (2006),
22. {\it Cuzzi et al.} (1984),
23. {\it Holland et al.} (1998),
24. {\it Wilner et al.} (2002),
25. {\it Su et al.} (2005),
26. {\it Greaves et al.} (2005),
27. {\it Marengo et al.} (2005),
28. {\it Wyatt et al.} (2005)
29. {\it Greaves et al.} (2004)
}
\end{deluxetable}

The scattered light colors are now known for six debris disks (see Table
1). In many of these an asymmetry factor (g) has also been measured;
larger grains are generally more forward-scattering. For disks in which
the mid-infrared emission has also been resolved, the amount of
scattered light compared to the mid-infrared emission from the same
physical areas enables a calculation of the albedo (albedo =
Qsca/(Qsca+Qabs)).  The albedo of canonical {\it Draine and Lee} (1984)
astronomical silicates is such that (for 0.5-1.6 $\mu$m observations),
grains smaller than 0.1 $\mu$m Rayleigh scatter and are blue, grains
larger than 2 $\mu$m scatter neutrally, and grains in between appear
slightly red.  In the case of a power law distribution of grain sizes,
such as that of a collisional cascade (equation 2), 
the scattering is dominated by the smallest
grains. Thus the colors in the Table have been explained by tuning the
smallest grain size to give the appropriate color.  Rarely has the
scattered color been modeled simultaneously with other constraints on
similar sized grains such as 8--13 $\mu$m spectra.
If observations of scattered light at longer
wavelengths continue to show red colors, the fine tuning of the minimum
grain size of astronomical silicates will fail to work. More realistic
grains may be porous aggregates where the voids may contain ice. Few
optical constants for these are currently available in the literature.

\subsection{Debris Disks Resolved in Sub--mm Emission}

Resolved observations from JCMT/SCUBA in the sub-millimeter at 850
$\mu$m by {\it Holland et al.}  (1998) and 
{\it Greaves et al.} (1998) led the way
in placing constraints on cold dust morphologies for four disks
(Fomalhaut, Vega, $\beta$ Pic, and $\epsilon$ Eri), including rings of dust
at Kuiper-belt like distances from stars and resolving clumps and inner
holes.  Since PPIV, higher spatial resolution images at 350 - 450
$\mu$m revealed additional asymmetries interpreted as indications for
planets ({\it Holland et al.}, 2003; {\it Greaves et al.}, 2005; 
{\it Marsh et al.}, 2005).
Perhaps most excitingly, the structure of the disk surrounding 
$\epsilon$ Eri appears to be
rotating about the star.  A longer time baseline for the motion of disk
clumps will reveal the mass and eccentricity of the planet responsible
for their generation ({\it Greaves et al.}, 2005). Finally, three additional
disks -- $\tau$ Ceti ({\it Greaves et al.}, 2004), HD 107146 
({\it Williams et al.}, 2004), and $\eta$ Corvi ({\it Wyatt et al.}, 2005), 
were resolved by JCMT/SCUBA.
Interferometric imaging of one debris disk, Vega, allowed the first
measurement of structure at a wavelength of 1 mm 
({\it Koerner et al.}, 2001; {\it Wilner et al.}, 2002). Again, the presence of
clumps could be explained by the influence of a planet 
({\it Wyatt}, 2003). 
It is interesting to note that three A-type stars, with
masses up to twice that of the Sun and luminosities up to tens of times
higher show dynamical evidence for planets.

\subsection{Debris Disks Resolved in IR Emission}

Ground-based 8~m class telescopes provide the best spatial resolution
for imaging disks, but are hampered by low sensitivity -- only two
debris disks ($\beta$ Pic and HR 4796) are definitively resolved at
12--25 $\mu$m from the ground.

Spitzer, with its ten times smaller aperture is able to resolve only nearby
disks. With MIPS, Spitzer has surprised observers with images of $\beta$ Pic,
$\epsilon$ Eri, Fomalhaut, and Vega that look quite different from their
submillimeter morphologies.  If Spitzer's sensitivity picked up the
Wien tail of the submillimeter grain emission or if the smaller mid-infrared
emitting grains were co-located with their larger progenitor bodies, then the
morphologies would be the same.  In the case of Fomalhaut, the MIPS 24 
$\mu$m 
flux originates in a Zodiacal-like region closer to the star {\it and} 
the planetesimal ring while the 70 $\mu$m flux does indeed trace the ring
({\it Stapelfeldt et al.}, 2004, and Fig. 3). As for the solar system, there may be
separate populations of planetesimals (analogous to the asteroid and Kuiper
belts) generating dust.

Surprisingly, however, the 24 and 70 $\mu$m images of
Vega actually have larger radii than the submillimeter ring or
millimeter clumps ({\it Su et al.}, 2005). This emission seems to trace small
grains ejected by radiation pressure.  Vega is only slightly more
luminous than Fomalhaut, so the minimum grain size generated in
collisions within the disk would have to finely tuned to below the
blowout size for Vega and above the blowout size for Fomalhaut 
for a unified disk model 
(see equation 3).  $\epsilon$ Eri looks about as expected with the
70 $\mu$m emission from the region of the submm ring ({\it Marengo et al.}, 2005).
An inner dust population might be expected if Poynting-Robertson drag
is important for the dust dynamics of this system (see Section 4).  The
absence of close-in dust may indicate that it is ejected by the
postulated planet.  It is also interesting that Spitzer did not resolve
any of the other nearby disks imaged including ones
resolved in the submm such as $\beta$ Leo 
(see however new results on
$\eta$ Corvi by {\it Bryden et al.}, in preparation).  
It is possible in these cases that the grain sizes are so 
large that Spitzer cannot see the Wien-side of such cold emission and/or
that their viewing geometries (nearly face-on) were unfavorable.

Spatially resolved spectroscopy has been obtained for only one debris
disk, $\beta$ Pic. These spectra provided information on collision
rates, with small silicate grains only observed within
20 AU of the star and thermal processing, with 
crystalline silicate fractions higher closer to the star
({\it Weinberger et al.}, 2003; {\it Okamoto et al.}, 2004). 
Of the stars in Table 1 with measured
scattered light, only $\beta$ Pic, HR 4796, and Fomalhaut have been
resolved in the infrared.

Only silicates with D $<$ 4$\mu$m show silicate emission.  In the
Zodiacal dust, this is only $\sim$ 10\% and the "typical" grain size is
100 $\mu$m ({\it Love and Brownlee}, 1993).  
Without resolving disks, the line-to-continuum ratio of
the mid-infrared silicate bands at 10--20 $\mu$m, which in principle
reflects the proportion of small grains, can be diluted by flux from
cold grains.  Many debris disks with 12$\mu$m excess show no silicate
emission ({\it Jura et al.}, 2004) with the implication that their grains are
larger than 10 $\mu$m.  The unfortunate consequence is that direct
compositional information is hard to acquire.

\begin{figure*}
\epsscale{1.5}
\plotone{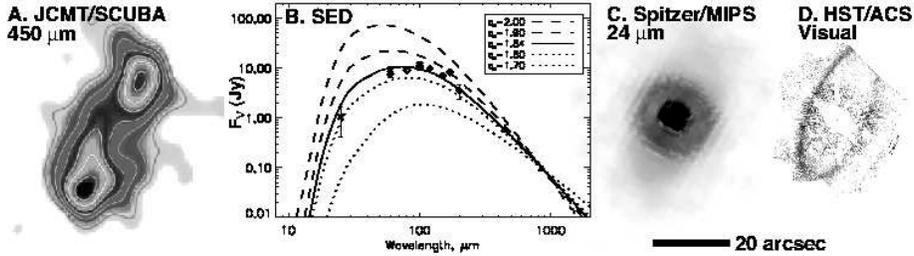}
\caption{\small The Fomalhaut disk is one of the few to have
been resolved in (A) the submillimeter ({\it Holland et al.}, 2003), 
(C) the thermal infrared ({\it Stapelfeldt et al.}, 2004), and
(D) scattered visual light ({\it Kalas et al.}, 2005). 
When only mid-infrared total fluxes and the
submillimeter images were available, 
{\it Wyatt and Dent} (2002) made models (B) 
using compact silicate grains. The addition of the mid-infrared images allows a
separation between warm (T$\sim$150 K) 
dust in an inner portion of the ring not seen
in the submm and the outer colder ring.  The addition of the scattered light
image allows a more accurate determination of the ring geometry including
a direct detection of the offset center of symmetry, similar to that observed
in HR 4796 ({\it Wyatt et al.}, 1999). 
In future work, the silicate model must be tuned 
to fit the dust scattered light (albedo) as well as emissivity.}
\end{figure*}

\subsection{Detections of Remnant Gas}

A useful definition of a debris disk is that it is gas free, because
then the dust dynamics are dominated by the processed described in
Section 4 unmodified by gas drag ({\it Takeuchi and Artymowicz}, 2001). 
However, debris
disks can have small amounts of gas released in the evaporation of
comets or destructive grain-grain collisions.
The most sensitive gas measurements are made with ultraviolet absorption
spectroscopy of electronic transitions. These transitions are 
strong and trace atomic and molecular gas at a wide range of
temperatures.  Yet since absorption spectroscopy probes only a
single line of sight, it is only very useful for edge-on disks and it
remains uncertain how to go from measured column densities to total disk
masses. 

The edge-on disks around the coeval $\beta$ Pic and AU Mic provide
strong constraints on the persistence of gas into the debris disk phase.
The total measured gas mass in $\beta$ Pic is $\rm 7\times 10^{-4}
M_{\oplus}$ while the upper limit (set by limits on HI) is 0.03 M$_{\oplus}$
({\it Roberge et al.}, 2006).  
Because the CO/H$_2$ ratio is more like that of
comets than of the ISM (CO is actually more abundant than H$_2$), the
gas is presumably second--generation just as the dust is 
({\it Lecavelier des Etangs et al.}, 2001).  In AU Mic, the
upper limit to the gas mass from the non-detection of molecular hydrogen
is 0.07 M$_{\oplus}$ ({\it Roberge et al.}, 2005). 
$\beta$ Pic and AU Mic differ in
luminosity by a factor of 90 but both were able to clear their
primordial gas in under 12 Myr.  Similar upper limits on the gas mass
are also observed for the slightly younger, slightly less edge-on disk
around HR 4796A ({\it Chen}, 2002).

Beyond total mass, a detailed look at the $\beta$ Pic disk reveals a
wide range of atomic species in absorption with an up-to-date inventory
given in {\it Roberge et al.} (2006). 
In addition, the spatial distribution of
gas in $\beta$ Pic is also imaged by long slit high spectral resolution
spectroscopy ({\it Brandeker et al.}, 2004). Atomic gas species such as sodium,
iron, and calcium are all distributed throughout the disk with Keplerian
line-of-sight velocities.  The observation that iron, which should 
experience strong radiation pressure and be ejected on orbital timescales,
has such low velocities remains a puzzle
({\it Lagrange et al.}, 1998).  At this time, the best explanation for why the gas
is not ejected by radiation pressure is that the ions strongly couple
via Coulomb forces enhanced by an overabundance of carbon gas ({\it Fernandez
et al.}, 2006; {\it Roberge et al.}, 2006).  Most of the gas in the disk is
ionized by a combination of stellar and interstellar UV.  Remaining
puzzles are the vertical distribution of calcium gas, which is actually
located predominantly away from the midplane ({\it Brandeker et al.}, 2004) and why
there exists such a large overabundance of carbon in the stable gas
({\it Roberge et al.}, 2006).

\section{Overview of Debris Disk Models}

\subsection{Basic Dust Disk Physics}

As described above, knowledge concerning general trends in 
the evolution of dust as a function of radius (see Section 2), 
as well as detailed information 
concerning particle composition and size distribution (see Section 3), 
abounds.  However, {\it understanding} these trends and placing specific 
systems in context requires that we interpret these data
in the context of robust physical theory.  Models of debris 
disks have to explain two main observations:
the radial location of the dust and its size distribution.
There are two competing physical processes that determine
how these distributions differ from that of the parent planetesimals
which are feeding the dust disk.

{\em Collisions} All material in the disk is subject to collisions with other objects,
both large and small.
If the collision is energetic enough, the target particle is destroyed
(a catastrophic collision)
and its mass redistributed into smaller particles.
Lower energy collisions result in cratering of the target particle
or accretion of the target and impactor.
It is catastrophic collisions that replenish the dust we see in
debris disks, and collisional processes are responsible for shaping
a disk's size distribution.

Both experimental ({\it Fujiwara et al.}, 1989) and numerical 
({\it Benz and Asphaug}, 1999)
work has been used to determine the specific incident energy required
to catastrophically destroy a particle, $Q_D^\star$.
This energy depends on particle composition, as well
as the relative velocity of the collision ($v_{rel}$), but
to a greater extent is dependent on the size of the target.
It is found to lie in the range $Q_D^\star=10^0-10^6$ J kg$^{-1}$,
which means that for collision velocities of $\sim 1$ km s$^{-1}$
particles are destroyed in collisions with other particles that are
at least $X=0.01-1$ times their own size, $D$.  
The collision velocity depends on the eccentricities and
inclinations of the particles' orbits, the mean values of which
may vary with particle size after formation (e.g., {\it Weidenschilling
et al.}, 1997). For planetesimal growth to occur both have to be
relatively low $\sim 10^{-3}$ to prevent net destruction of particles.
However to initiate a collisional cascade something must have
excited the velocity dispersion in the disk to allow collisions to
be catastrophic. Models which follow the collisional evolution of
planetesimal belts from their growth phase through to their cascade phase
show that this switch may occur after the formation of a planet
sized object ({\it Kenyon and Bromley}, 2002a, 2004) or from excitation by a
passing star ({\it Kenyon and Bromley}, 2002b). 

A particle's collisional lifetime is the mean time between catastrophic
collisions.
This can be worked out from the catastrophic collision rate which is
the product of the relative velocity of collisions and the
volume density of cross-sectional area of the impactors larger
than $XD$.
For the smallest particles in the distribution, for which collisions
with any other member of the distribution is catastrophic, their
collisional lifetime is given by:
\begin{equation}
  t_{coll} = t_{per}/4\pi \tau_{eff},
  \label{eq:tcoll}
\end{equation}
where $t_{per}$ is the orbital period and $\tau_{eff}$ is the surface
density of cross-sectional area in the disk which when 
multiplied by the absorption efficiency of the grains
gives the disk's face--on optical depth ({\it Wyatt and Dent}, 2002). 
Larger particles have longer collisional lifetimes.

In an infinite collisional cascade in which the outcome of
collisions is self-similar (in that the size distribution of collision
fragments is independent of the target size), collisions are expected
to result in a size distribution with
\begin{equation}
  n(D) \propto D^{-3.5}
  \label{eq:nd}
\end{equation}
({\it Dohnanyi}, 1969; {\it Tanaka et al.}, 1996).
Such a distribution has most of its mass in the largest planetesimals,
but most of its cross-sectional area in the smallest particles.

{\em Radiation pressure and P-R drag} Small grains are affected by their interaction with stellar radiation
which causes a force on the grains which is parameterized by
$\beta$, the ratio of the radiation force to stellar gravity.
This parameter depends on the size of the grain, and to a lesser
extent on its composition.
For large particles $\beta$ can be approximated by
\begin{equation}
  \beta = (0.4 \mu m/D)(2.7g cm^{-3}/\rho)
          (L_\star/M_\star),
  \label{eq:beta}
\end{equation}
where $\rho$ is the grain density and $L_\star$ and $M_\star$ are
in units of $L_\odot$ and $M_\odot$ ({\it Burns et al.}, 1979).
However, this relation breaks down for particles comparable in size
to the wavelength of stellar radiation for which a value of $\beta$
is reached which is independent of particle size ({\it Gustafson}, 1994).

The radial component of the radiation force is known as radiation pressure.
For grains with $\beta>0.5$ (or $D < D_{bl}$), which corresponds to
sub-micron sized grains near a Sun-like star, radiation pressure
causes the grains to be blown out of the system on hyperbolic
trajectories as soon as they are created.
Since grains with $\beta=1$ have no force acting on them, the blow-out
timescale can be estimated from the orbital period of the parent
planetesimal:
\begin{equation}
  t_{bl} = \sqrt{a^3/M_\star},
  \label{eq:tbl}
\end{equation}
where $a$ is the semimajor axis of the parent in AU, 
and $t_{bl}$ is the
time to go from a radial distance of $a$ to $6.4a$.
In the absence of any further interaction, such grains have a surface
density distribution that falls off $\propto r^{-1}$.

The tangential component of the radiation force in known as
Poynting-Robertson (P-R) drag.
This acts on all grains and causes their orbits to decay in to the star
(where the grains evaporate) at a rate $\dot{a} = -2\alpha/a$, where
$\alpha = 6.24 \times 10^{-4}M_\star/\beta$.
Thus the evolution from $a$ to the star takes
\begin{equation}
  t_{pr} = 400(a^2/M_\star)/\beta
  \label{eq:tpr}
\end{equation}
in years.
In the absence of any further interaction, such grains have a surface
density distribution that is constant with the distance from the star.

{\em Other processes} Other physical processes acting on dust in debris disks range from gas
drag to stellar wind drag, Lorentz forces on charged grains and
sublimation.
Many of these have been determined to be unimportant in the physical
regimes of debris disks.
However, it is becoming clear that for dust around M stars the force of the
stellar wind is important both for its drag component
({\it Plavchan et al.}, 2005) and its pressure component 
({\it Strubbe and Chiang}, in press). 
Gas drag may also be important in young debris disks.
While the quantity of gas present is still poorly known, 
if the gas disk is sufficiently dense then gas drag can significantly
alter the orbital evolution of dust grains.
This can result in grains migrating to different radial locations
from where they were created, with different sizes ending up at
different locations (e.g., {\it Takeuchi and Artymowicz}, 2001; 
{\it Klahr and Lin}, 2001; {\it Ardila et al.}, 2005; 
{\it Takeuchi et al.}, 2005).  For 
$\beta$ Pic it has been estimated that gas drag becomes important
when the gas to dust ratio exceeds 1 ({\it Th\'{e}bault and Augereau}, 2005).

\subsection{Model Regimes}

A debris disk that is not subjected to the stochastic mass--loss
processes discussed in Sections 4.4 and 5, will evolve in steady--state 
losing mass through radiation processes acting on small grains: 
through P-R drag and consequently evaporation close to the star,
or through collisional grinding down and consequently
blow-out by radiation pressure.
The competition between collisions and P-R drag was explored
in {\it Wyatt} (2005) which modeled the dust distribution expected if
a planetesimal belt at $r_0$ is creating dust of just one size (see
Fig. 4). 
The resulting distribution depends only on the parameter $\eta_0 =
t_{pr}/t_{coll}$.
If the disk is dense ($\eta_0 \gg 1$), then collisions occur much faster
than P-R drag and the dust remains confined to the planetesimal belt,
whereas if the disk is tenuous ($\eta_0 \ll 1$) then the dust suffers
no collisions before reaching the star and the dust distribution is
flat as expected by P-R drag.
While this is a simplification of the processes going on in debris
disks, which are creating dust of a range of sizes, it serves to
illustrate the fact that disks operate in one of two regimes:
collisional or P-R drag dominated.
These regimes are discussed in more detail below.

\begin{figure}
 \epsscale{0.75}
 \plotone{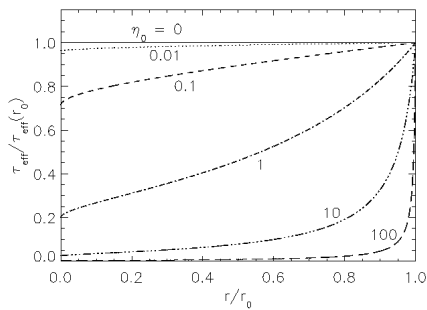}
 \caption{\small Surface density distribution of dust
  created in a planetesimal belt at $r_0$ which evolves
  due to collisions (which remove dust) and P-R drag
  (which brings it closer to the star) ({\it Wyatt}, 2005).
  Assuming the dust is all of the same size, the resulting
  distribution depends only on $\eta_0$, the ratio of the
  collisional lifetime to that of P-R drag.}  
\end{figure}

{\em Collisionally dominated disks} In a collisionally dominated disk ($\eta_0 \gg 1$) it is possible to
ignore P-R drag, since the  cumulative migration of particles over
all generations from planetesimal to $\mu$m-sized grain is negligible
(e.g., {\it Wyatt et al.}, 1999).
This is because P-R drag lifetimes increase $\propto D$, whereas
collisional lifetimes increase $\propto D^{0.5}$ (assuming the
distribution of eq.~\ref{eq:nd}) meaning that the migration undergone
before a collision becomes vanishingly small for large
particles.

There are two components to a collisionally dominated
debris disk: dynamically bound grains at the same radial location
as the planetesimals, and unbound grains with an $r^{-1}$ distribution
beyond that.
The short lifetime of the unbound grains (eq.~\ref{eq:tbl}) suggests
that their number density should be extremely tenuous, and should fall
below that expected from an extrapolation of the collisional cascade
distribution.
However, recent observations indicate that in some imaged
debris disks they are being replenished at a rate sufficient for these
grains to dominate certain observations (e.g., {\it Telesco et al.}, 2000;
{\it Augereau et al.}, 2001; {\it Su et al.}, 2005), implying a comparable
cross-sectional area in these particles to that in bound grains
as currently observed. 

The size distribution in a collisionally dominated disk varies
somewhat from that given in eq.~\ref{eq:nd}, since that assumes
an infinite collisional cascade.
If the number of blow-out grains falls below that of the
collisional cascade distribution, then since these particles would be
expected to destroy particles just larger than themselves, their low
number causes an increase in the equilibrium number of particles just
above the blow-out limit.
This in turn reduces the equilibrium number of slightly larger
particles, and so on;
i.e., this causes a wave in the size distribution which continues
up to larger sizes ({\it Th\'{e}bault et al.}, 2003).
If, on the other hand larger quantities of blow-out grains are
present (e.g., because their number is enhanced by those driven out
from closer to the star), then this can actually reduce the equilibrium
number of particles just above the blow-out limit ({\it Krivov et al.}, 2000).

The long term evolution of a collisionally dominated disk was considered
by {\it Dominik and Decin} (2003).
They considered the case where the dust disk is fed by planetesimals of a
given size, $D_c$, and showed how collisions
cause the number of those planetesimals, $N_c$, to follow:
\begin{equation}
  N_c(t) = N_c(0)/[1+2t/t_c(0)],
\end{equation}
where $t_c$ is the collisional lifetime of the colliding planetesimals
at $t=0$.
In other words, the evolution is flat until the disk is old
enough for the majority of the planetesimals to have collided with each other
(i.e., when $t > t_c$),
thus eroding their population, at which point their number falls off
$\propto t^{-1}$.
Since the size distribution connecting the dust to the number of
planetesimals is given by eq.~\ref{eq:nd}, it follows that the
cross-sectional area of emitting dust has the same flat or $t^{-1}$
evolution as does the total mass of material in the disk which is
dominated by planetesimals of size $D_c$.
{\it Dominik and Decin} (2003) also noted ways of changing the evolution,
e.g., by introducing stirring.

The quantity of blow-out grains in the disk does not follow the same
evolution, since their number is determined by the equilibrium between
the rate at which the grains are created and that at which they are
lost (eq.~\ref{eq:tbl}).
The rate at which they are created depends on details of the physics
of collisions, but since the rate at which dust is produced by
planetesimals is $\propto N_c^2$, it follows that their population
should fall off $\propto t^0$ or $t^{-2}$ depending on whether
$t<t_c$ or $t>t_c$.

{\em P-R drag dominated disks} A conclusion shared by {\it Dominik and Decin} (2003) and 
{\it Wyatt} (2005) is
that none of the debris disks detected with current instrumentation
is in the P-R drag dominated regime.
{\it Wyatt} (2005) explained this as a consequence of the fact that such
disks are of too low mass for their emission to be comparable to
that of the stellar photosphere.
Thus the detection of such disks requires calibration to a few \%
in the mid- to far-IR, or discovery in the sub-mm.
However, the zodiacal cloud (and presumably dust from the Kuiper
belt) in the solar system is a good example
of a P-R drag dominated disk.

It is not possible to completely ignore collisions in a P-R drag dominated
disk, since, while the smallest dust makes it to the star without suffering
a collision, the largest grains are in a collisionally dominated regime,
with intermediate sizes having distributions closer to that of $\eta_0=1$
in Fig. 4). 
Matters are complicated by the way P-R drag affects the size distribution.
If collisional processes in a planetesimal belt are assumed to create
dust at a rate that results in the size distribution of
eq.~\ref{eq:nd} in the planetesimal belt, then since small dust
migrates faster than small dust (eqs.~\ref{eq:beta} and \ref{eq:tpr})
then the size distribution of the dust affected by P-R drag should
follow
\begin{equation}
  n(D) \propto D^{-2.5} \label{eq:ndpr}
\end{equation}
({\it Wyatt et al.}, 1999), a distribution in which most of the cross-sectional
area is in the largest particles in that distribution.
In other words, the cross-sectional area should be dominated by grains for
which P-R drag and collisional lifetimes are roughly equal, with that
size varying with distance from the planetesimal belt.
This reasoning is in agreement with observations of the size distribution
of interplanetary dust in the vicinity of the Earth 
({\it Love and Brownlee}, 1993;
{\it Wyatt et al.}, 1999; {\it Grogan et al.}, 2001).
{\it Dominik and Decin} (2003) also looked at 
the evolution of P-R drag dominated disks within the model described above.
They concluded that the quantity of visible grains should fall off
$\propto t^{-2}$.

\subsection{Formation of inner hole}

Perhaps the most important discovery about debris disks is the fact that
there are inner holes in their dust distribution.
It is often suggested that planet-sized bodies are required interior to
the inner edge of the debris disk to maintain the inner holes, because
otherwise the dust would migrate inward due to P-R drag thus filling in
the central cavity ({\it Roques et al.}, 1994).
It is certainly true that a planet could maintain an inner hole by a
combination of trapping the dust in its resonances ({\it Liou and Zook}, 
1999), scattering the dust outward ({\it Moro-Martin and Malhotra}, 2002),
and accreting the dust ({\it Wyatt et al.}, 1999).
However, a planet is not required to prevent P-R drag filling in the holes
in the detected debris disks, since in dense enough disks collisional grinding down already
renders P-R drag insignificant ({\it Wyatt}, 2005).

What the inner holes do require, however, is a lack of colliding
planetesimals in this region.
One possible reason for the lack of planetesimals close to the star
is that they have already formed into planet-sized objects, since
planet formation processes proceed much faster closer to the star
({\it Kenyon and Bromley}, 2002).
Any remaining planetesimals would then be scattered out of the
system by chaos induced by perturbations from these larger bodies
(e.g., {\it Wisdom}, 1980).

\subsection{Steady-state vs Stochastic Evolution}

Much of our understanding of debris disks stems from our
understanding of the evolution of the zodiacal cloud.
This was originally assumed to be in a quasi steady-state.
However, models of the collisional evolution of the asteroid
belt, and the dust produced therein, showed significant peaks
in dust density occur when large asteroids collide releasing
quantities of dust sufficient to affect to total dust content
in the inner solar system ({\it Dermott et al.}, 2002).
Further evidence for the stochastic evolution of the asteroid belt
came from the identification of asteroid families created in the
recent (last few Myr) break-up of large asteroids ({\it Nesvorn\'{y} et
al.}, 2003).
The link of those young families to the dust band features in
the zodiacal cloud structure ({\it Dermott et al.}, 2002) and to peaks in the
accretion rate of $^3$He by the Earth ({\it Farley et al.}, 2005) 
confirmed the stochastic 
nature of the inner solar system dust content,
at least on timescales of several Myr.
More recently the stochastic nature of the evolution of debris
disks around A stars 
has been proposed by {\it Rieke et al.} (2005) based on the
dispersion of observed disk luminosities. 
Several debris disks are observed to have small grains 
(with very short lifetimes) at radii inconsistent with  
steady-state configurations over the lifetime of the star (e.g.,
{\it Telesco et al.}, 2005; {\it Song et al.}, 2005; {\it Su et al.}, 2005).

The arguments described previously considered the steady-state
evolution of dust created in a planetesimal belt at single radius.
The same ideas are still more generally applicable to stochastic
models, since a situation of quasi-steady state is reached relatively
quickly at least for small dust for which radiation and collision
processes balance on timescales of order 1 Myr (depending on disk
mass and radius).

Stochastic evolution of the type seen in the zodiacal cloud arises
from the random input of dust from the destruction of large planetesimals.
Whether it is possible to witness the outcome of such events in
extrasolar debris disks is still debated for individual objects.
This is unlikely to be the case for dust seen in the sub-mm,
since the large dust mass observed requires a collision between
two large planetesimals ($>1400$ km for dust seen in Fomalhaut),
and while such events may occur, the expected number of such objects
makes witnessing such an event improbable ({\it Wyatt and Dent}, 2002).
Observations at shorter wavelengths (and closer
to the star) probe lower dust masses, however, and these observations
may be sensitive to detecting such events 
({\it Telesco et al.}, 2005; {\it Kenyon and Bromley}, 2005).

Debris disk evolution may also be affected by external influences.
One such influence could be stirring of the disk by stars which pass
by close to the disk ({\it Larwood and Kalas}, 2001; 
{\it Kenyon and Bromley}, 2002b).
However, the low frequency of close encounters with field stars means
this cannot account for the enhanced dust flux of all debris disk
candidates, although such events may be common
in the early evolution of a disk when it is still in a dense cluster
environment.
Another external influence could be the passage of the disk through a
dense patch of interstellar material which either replenishes the
circumstellar environment with dust or erodes an extant, but low
density debris disk 
({\it Lissauer and Griffith}, 1989; {\it Whitmire et al.}, 1992).

Other explanations which have been proposed to explain sudden
increases in dust flux include the sublimation of supercomets
scattered in close to the star ({\it Beichman et al.}, 2005).

It is also becoming evident that the orbits of the giant planets
have not remained stationary over the age of the solar system
({\it Malhotra}, 1993; {\it Gomes et al.}, 2005).
The recently investigated stochastic 
component of giant planet orbital evolution can
explain many of the features of the solar system including the 
period of Late Heavy Bombardment (LHB) which rapidly depleted the 
asteroid and Kuiper belts, leading to enhanced collision rates in 
the inner solar system. 
Such an event in an extrasolar system would dramatically increase
its dust flux for a short period, but would likely do so only once
in the system's lifetime.
A similar scenario was also proposed by {\it Thommes et al.} (1999)
to explain the LHB wherein the giant cores that formed between Jupiter
and Saturn were thrown outwards into the Kuiper Belt by chaos at a late
time.
Again this would result in a spike in the dust content of an extrasolar
system.  These ideas are applied to own solar system in the next section. 

\section{Comparison to our Solar System} 

Our asteroid belt (AB) and Kuiper-Edgeworth belt (KB) contain planetesimals
that accreted during the earliest epochs of the solar system's formation,
plus fragments from subsequent collisions (e.g., {\it Bottke et al.}, 2005; 
{\it Stern and Colwell}, 1997). 
Collisions in both belts should generate populations of dust
grains analogous to extrasolar debris disks.  The dust population extending
from the AB is directly observed as the zodiacal cloud, whereas dust associated
with the KB is as yet only inferred ({\it Landgraf et al.}, 2002).   An observer
located 30 pc from the present solar system would receive approximately 
70 microJy at 24 $\mu$m and 20 microJy at 70 $\mu$m from the AB plus zodi cloud,
in contrast to 40 milliJy, hereafter mJy, 
(24 $\mu$m) and 5 mJy (70 $\mu$m) from the Sun.
The luminosity of the KB dust component is less certain but flux densities 
from 30 pc of about 0.4 mJy at 24 $\mu$m and 4 mJy at 70 $\mu$m correspond to
an estimated KB planetesimal mass of 0.1 M$_{\oplus}$ (see below for details of these
calculations).

The solar system's original disk contained much more solid mass in the AB and KB
zones than at present.  A minimum-mass solar nebula would have had 3.6 M$_{\oplus}$ of
refractory material in the primordial AB between r = 2.0 and 4.0 AU whereas 
now the AB contains only $5 \times 10^{-4}$ M$_{\oplus}$, and only $2 \times 10^{-4}$ M$_{\oplus}$ if
the largest object Ceres is excluded.  
In contrast, the masses of
Earth and Venus are close to the minimum-mass 
nebular values for their respective 
accretion zones.  Likewise, the primordial KB must have had 10-30 M$_{\oplus}$ so that the
observed population of large objects could have formed in less than $10^8$ years
before gravitational influence of the planets made further accretion impossible
({\it Stern}, 1996).  
The present KB contains no more than a few $\times 0.1$ M$_{\oplus}$ based
on discovery statistics of massive objects (discussed by 
{\it Levison and Morbidelli}, 2003) 
and upper limits to IR surface brightness of collisionally evolved dust
({\it Backman et al.}, 1995; {\it Teplitz et al.}, 1999).

How did the missing AB and KB masses disappear?  It is unlikely that purely
``internal" collisional processes followed by radiation pressure-driven removal
of small fragments is responsible for depletion of either belt.   Persistence
of basaltic lava flows on Vesta's crust is evidence that the AB contained no
more than 0.1 M$_{\oplus}$ 6 Myr after the first chondrules formed ({\it Davis et al.}, 
1985; {\it Bottke et al.}, 2005).  This implies a
factor of at least 40x depletion of the AB zone's mass by that time, impossible
for purely collisional evolution of the original amount of material (reviewed by
{\it Bottke et al.}, 2005; cf. Section 4 of this chapter).  Also, the present AB
collisional ``pseudo-age", i.e.\ the model time scale for a purely
self-colliding AB to reach its present density, is of order twice the 
current age of the Solar System.  This is
further indication that the AB's history includes significant depletion
processes other than comminution.  Proposed depletion mechanisms include
sweeping of secular resonances through the AB as the protoplanetary disk's gas
dispersed ({\it Nagasawa et al.}, 2005)
and as Jupiter formed and perhaps migrated during the solar system's
first 10 Myr or so ({\it Bottke et al.}, 2005).

Similarly, several investigators have concluded that the primordial KB was
depleted by outward migration of Neptune that swept secular resonances through
the planetesimal population, tossing most of the small objects either inward
to encounter the other planets or outward into the KB's ``scattered disk".
That scenario neatly explains several features of the present KB in addition to
the mass depletion ({\it Levison and Morbidelli}, 2003; 
{\it Gomes et al.}, 2005; see the chapter by {\it Levison et al.}). 
This substantial re-organization of the solar system could
have waited a surprisingly long time, as much as 1.0 Gyr, driven by slow
evolution of the giant planets' orbits before becoming chaotic ({\it Gomes et al.}, 
2005).  The timing is consistent with the epoch of the Late Heavy Bombardment
(LHB) discerned in lunar cratering record.  Furthermore, 
{\it Strom et al.} (2005) point
out that, because Jupiter should have migrated inward as part of the same
process driving Neptune outward, the AB could have been decimated (perhaps
for the second time) at the same late era as the KB.

The simple model employed herein to track the history of the solar system's
IR SED involves calculating the collisional evolution of the AB and KB.  Each
belt is divided into 10 radial annuli that evolve independently.  At each time
step (generally set to 10$^6$ years) for each annulus is calculated: (1) the
number of parent body collisions, (2) the fragment mass produced and subtracted
from the parent body reservoir, and (3) mass lost via ``blowout" of the smallest
particles plus P-R drift from the belt inward toward the Sun.  Parent body
collisions are considered only statistically so the model has no capacity to
represent ``spikes" from occasional large collisions as discussed in 
Section 4.4.  Mass in grains that would
be rapidly ejected via radiation pressure ``blowout" is removed from the model
instantaneously when created.  If the collision timescale for bound grains of a
given size and location is shorter than the P--R removal time,
those grains are not allowed to drift interior to the belt 
and contribute to the inner zodiacal
cloud.  Thus, based on the theory explained in the previous section, if the
belt fragment density is above a certain threshold, the net mass loss is only
outward via direct ejection, not inward.  The terrestrial planets are not 
considered as barriers to P-R drift but Neptune is assumed able to consume or
deflect all grains, so the model dust surface density is set to zero between
4 and 30 AU.  The system SED is calculated using
generic grain emissivity that depends only on particle size. 
An indication that the model works well is
that it naturally predicts the observed zodiacal dust density as the output
from the observed AB large-body mass and spatial distributions without 
fine-tuning.

Our simple results agree with {\it Bottke et al.} (2005) 
and others' conclusion that
the AB and KB must both have been subject to depletions by factors of 10-100
sometime during their histories because simple collisional evolution would not 
produce the low-mass belts we see today.  The present AB and KB masses cannot
be produced from the likely starting masses without either (A) an arbitrary
continuous removal of parent bodies with an exponential time scale for both
belts of order 2 Gyr, shown in Fig 5 (top), or
(B) one sudden depletion event shown in Fig 5 (bottom), which involves
collisionally evolving the starting mass for 0.5 Gyr, then reducing each belt
mass by amounts necessary to allow collisional evolution to resume and continue
over the next 4.0 Gyr to reach the observed low masses of the two belts.

\begin{figure*}
\epsscale{2.0}
\plottwo{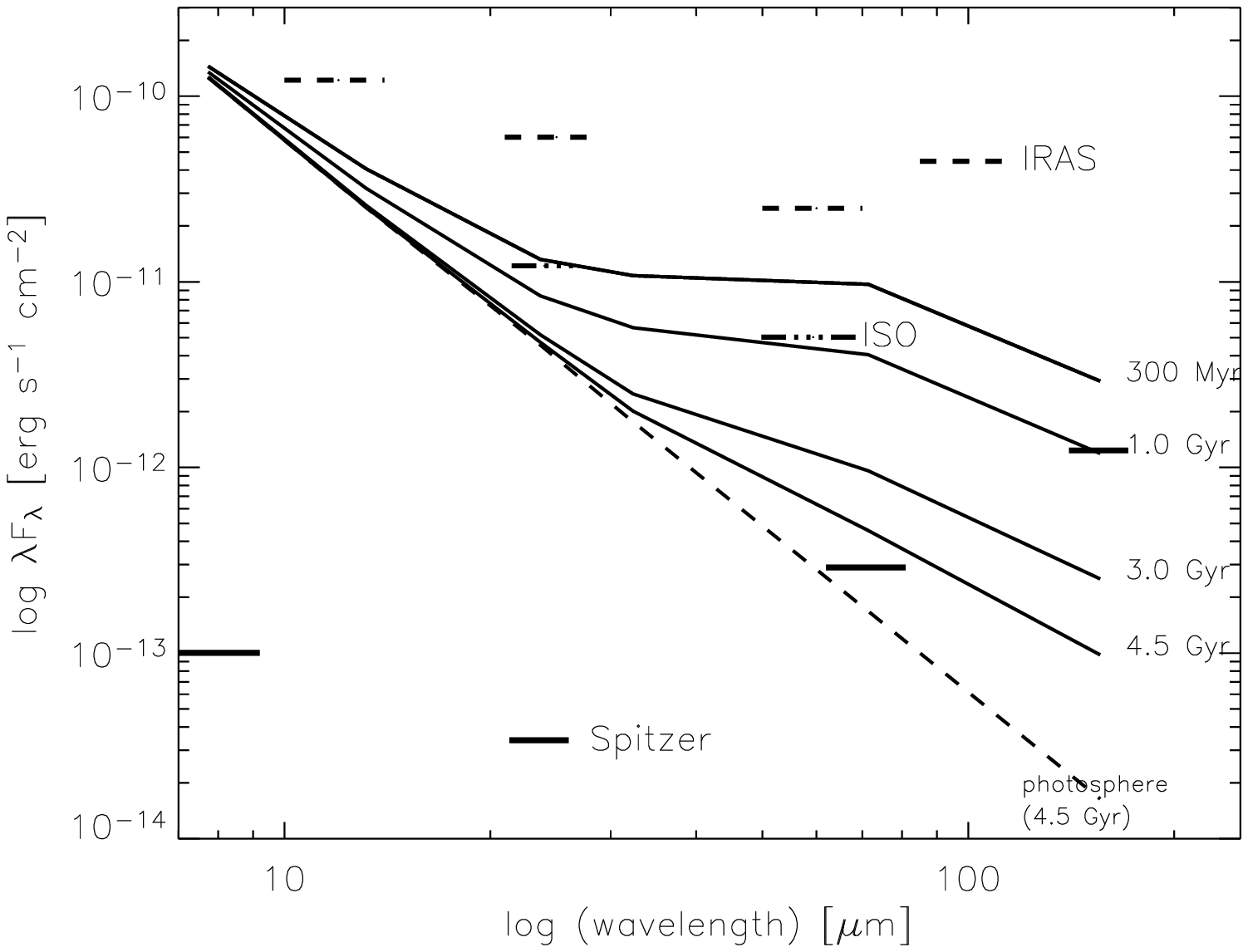}{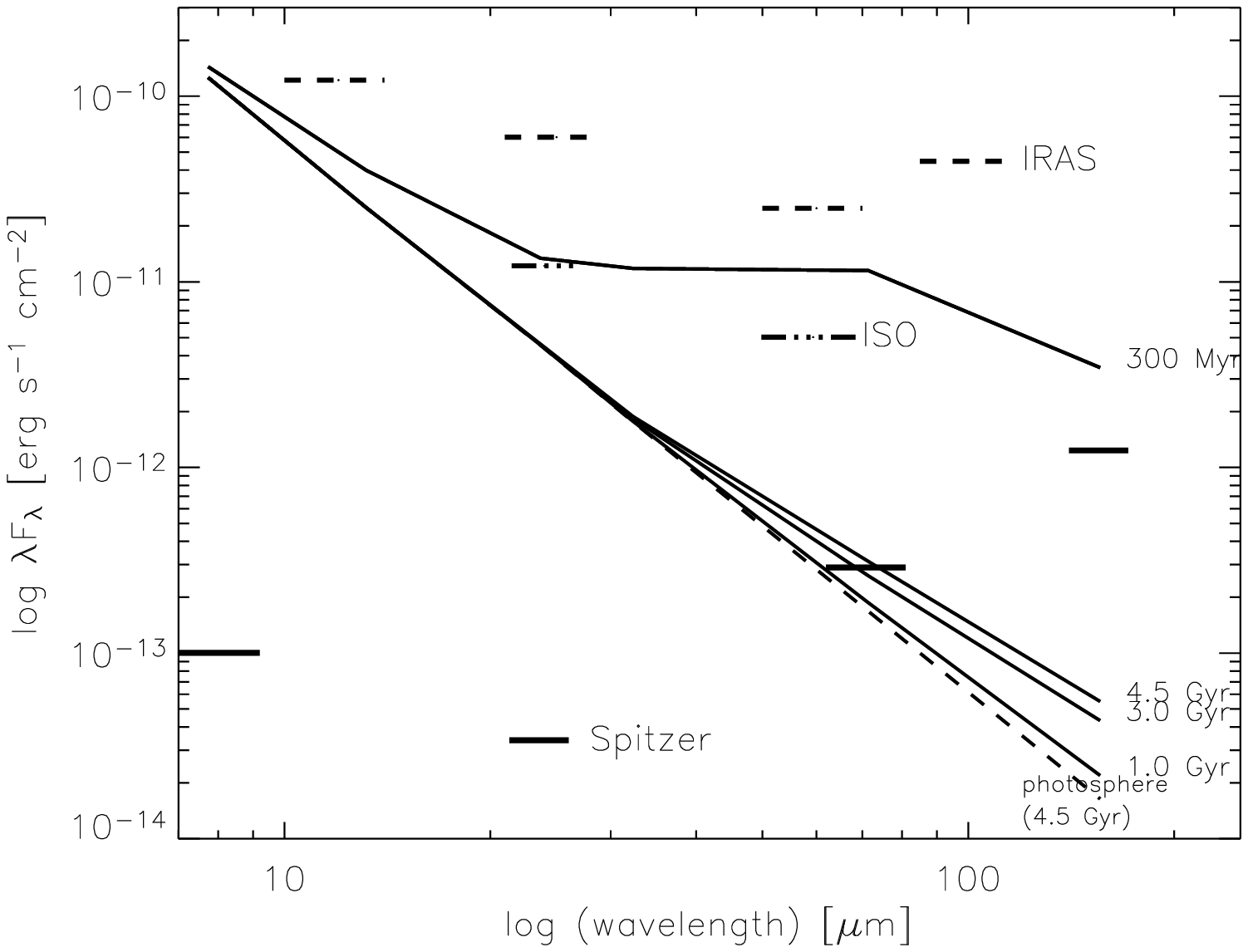}
\caption{\small Toy models for the evolution of the solar system spectral 
energy distribution from an age of 300 Myr to 4.5 Gyr: (A) with no late 
heavy bombardment shown (left) and (B) including the LHB (right)
 as discussed in the text. 
The long wavelength excess in (B) grows with time after the LHB because the 
Kuiper Belt has not yet reached equilibrium between dust production and 
dust removal.  Also shown are the 3 $\sigma$ 
sensitivity limits of IRAS, ISO, and Spitzer for a sun--like star
at a distance of 30 parsecs for comparison. 
}
 \end{figure*}

The toy model scenario A simply predicts that a planetary system
would have significant 10-30 $\mu$m flux up to an age
of 1 Gyr.   The general lack of observed mid-IR excesses in Spitzer
targets older than 30 Myr could mean:  a) most systems do not have
belts at temperatures like our asteroid belt, or b) most have
LHB-like events earlier in their histories.
A corollary of the present depleted AB having a large-body collision time scale
of 10 Gyr is that the AB/zodi system is nearly constant in time (e.g., 
equation 6).  Thus, extrapolating backward 
by {\it Dominik and Decin} (2003) t$^{-1}$ or t$^{-2}$
scaling laws is inappropriate (cf. Fig 1).  
Our model predicts that the AB/zodi and KB IR
luminosities both would only decrease by about 30 \% in 4.0 Gyr after 
the LHB or equivalent major clearing.   This agrees with the lunar cratering
record indicating that the AB, the earth-crossing asteroid population, and the
zodiacal  cloud have had nearly constant density for at least the past 3.0 Gyr.

In another chapter, {\it Levison et al.} discuss the idea that our planetary
system had a traumatic re-organization about 700 Myr after its formation.
An intriguing extension of this idea is that extrasolar debris disk systems
that seem brighter than their age cohorts may represent LHB-like events, i.e.\
collisions of small bodies excited by planet migration that can occur late in
a system's development at a time determined 
by details of the original planetary 
system architecture.   Our ``toy" model results compared with Spitzer
observations (e.g., {\it Kim et al.}, 2005) support
a picture in which many systems evolve according to the principles
articulated by {\it Dominik and Decin} (2003) unless
interrupted by an LHB event that might occur almost any time in the system's
history.  
After the LHB event the system is nearly clear of planetesimals and
dust and evolves very slowly.  It remains to be seen whether observations
with Spitzer can distinguish between hypotheses such as single
super-collisions or late episodes of debris belt clearing. 

\section{Summary and Implications for Future Work} 

Based on the discussions presented above, it is clear that the question
of how common solar systems like our own might be, depends in part on 
what radius in the disk one looks and at what age the comparison is made. 
We summarize our main results as follows: 
1) Warm circumstellar material inside of 1 AU dissipates
rapidly on timescales comparable to the cessation of accretion; 
2) The gas content of disks much older than 10 Myr is incapable
of forming giant planets; 3) While massive analogues to our
asteroid belt lacking outer disks appear to be rare overall (1--3 \%), 
warm disks (lacking inner hot dust) seem to enjoy a preferred epoch
around stars with ages between 10--300 Myr old; 4) 
Cold outer disks (analogous to our own Kuiper Belt, but much more massive) 
are found around 10--20 \% of sun--like stars; 5) Resolved
images of disks are crucial in order to remove degeneracies in 
debris disk modeling from SEDs alone; 6) Most debris disks 
observed to date are collisionally dominated dust systems and do 
not require the dynamical action of planets to maintain the
observed inner holes; 7) At least some disks are observed in a short--lived
phase of evolution and are not examples of the most
massive debris disks; and 8) Comparing the ensemble of observations
of disks surrounding other stars as a function of age 
to the evolution of our solar system 
requires detailed understanding of its dynamical evolution 
including the late--heavy bombardment era. 
Yet in affecting these comparisons, we must remember that we
do not yet have the sensitivity to observe tenuous debris disks comparable 
to our own asteroid belt or our Kuiper Belt.  

It is unclear whether debris systems significantly more massive 
(and therefore more easily detectable) than our own represent
a more or less favorable condition for planet formation.  
It may be that systems with planets might arise from disks with 
higher mass surface density and thus stronger debris signatures
at early times than disks lacking planets. However, if events
comparable to the dynamical re--arrangement of our solar system
(perhaps related to the lunar late--heavy bombardment) 
are common in planetary systems 
within the first few hundred million years we might
expect that debris disks lacking planets might be brighter
than those with planets at late times.  
{\it Beichman et al.} (2005a) 
present preliminary evidence that there may be some
connection between the presence of a massive debris disk and a
radial velocity planet within 5 AU.  It is interesting
to note that extrapolations of the detection frequency of 
extra--solar planets as a function of radius beyond current 
survey limits (see chapter by {\it Udry et al.}) 
suggest a frequency of extra--solar giant 
planets $>$ 1 M$_{JUP}$ $\sim$ 10--20 \% out to 20 AU, 
consistent with our debris disk statistics for G stars. 

How do results on debris disks compare as a function
of stellar mass? 
On theoretical grounds, one can argue that the mass of 
a circumstellar disk should not exceed $\sim$ 10--25 \% 
the mass of the central star ({\it Shu et al.}, 1990).  
Indeed {\it Natta et al.} (2000) presents evidence that
the disk masses around early type pre--main sequence
stars are more massive than their lower mass T Tauri 
counter--parts.  {\it Muzerolle et al.} (2003) also show
that disk accretion rates appear to correlate with 
stellar mass.  Historically, debris disks have been 
more commonly associated with A stars rather than 
G or M stars, but that has been largely attributable
to a selection effect:  it is easier to see smaller
amounts of dust surrounding higher luminosity 
objects in flux--limited surveys.  {\it Rieke et al.}
(2005) present evidence for a diminution in the frequency of mid--IR
excess emission surrounding A stars over 100--300 Myr. 
Their data indicate that over and above an evelope of
decay consistent with a t$^{-1}$ fall off, several 
objects show evidence for greater dust generation
rates consistent with their interpretation of stochastic
processes in planetesimal disks (see Sections 3 and 4 above). 
In general, the overall picture of A star debris disk 
evolution is remarkably consistent with that presented 
for sun--like stars suggesting that stellar mass does
not play a defining role in debris disk evolution. 
In contrast, primordial disks around higher mass stars
are more massive, and have shorter lifetimes 
({\it Hillenbrand et al.}, 1998; {\it Lada et al.}, in press), 
than disks around lower mass stars. 

{\it Greaves et al.} (2003) also 
present evidence from ISO observations concerning
the frequency of debris disks as a function of mass. 
They find that debris surrounding A stars is more 
common than around G stars, even for stars of the same 
age (though the observations were sensitive to different
amounts of debris as a function of stellar luminosity).  
They suggest that the difference is due to characteristic
lifetimes of debris becoming an increasing fraction of 
of the main sequence lifetime for higher mass
(shorter lived) stars, possibly because disk mass correlates
with star mass. 
{\it Plavchan et al.} (2005) present a survey for warm inner
debris surrounding young M dwarfs.  They explain their
lack of detections, which is contrary to expectations
from the timescale for P--R drag as a function of 
stellar luminosity, due to the effects of an enhanced
particulate wind from late--type stars compared to 
early--type stars. Yet, it is clear from recent work on 
low mass stars and brown dwarfs that they too possess 
primordial circumstellar disks 
when they are young (see chapter by {\it Luhman et al.}; 
{\it Apai et al.}, 2005) however their evolutionary properties
are as yet unclear.  Spitzer studies of debris disks
surrounding low mass stars and brown dwarfs at longer wavelengths 
are now underway.  
Combining data on A stars, G dwarfs, and M dwarfs, there
is, to date, no evidence for wildly divergent 
evolutionary histories for debris disks 
as a function of stellar mass averaged over
main sequence lifetimes.  Observed differences to date
can be explained in part by differences in 
dust mass upper limits as a function of 
stellar luminosity and 
assuming that the typical star to initial disk mass is
roughly constant. 

It is important to 
remember that most {\it sun--like} stars in the disk of 
Milky Way are binary ({\it Duquennoy and Mayor}, 1991), while
the binary fraction of low mass stars and brown dwarfs
may be lower (see chapter by {\it Burgasser et al.}). 
It is clear that the evolution of disks in the 
pre--main sequence phase can be influenced by
the presence or absence of a companion (see chapter 
by {\it Monin et al.}; {\it Jensen et al.}, 1996). 
Preliminary results
from Spitzer suggest that debris disk evolution
is not a strong function of multiplicity, 
and may even be enhanced in close binaries 
({\it Trilling et al.}, in preparation). 

What are the implications for the formation of 
terrestrial planets in disks surrounding 
stars of all masses in the disk of the Milky Way? 
We know that primordial accretion disks commonly 
surround very young stars (approaching 100 \%), 
and that gas--rich disks around more (less) massive 
stars are bigger (smaller), but last shorter (longer) amounts of time.   
Because of the surface density of solids in the disk, more massive
disks surrounding higher mass stars will probably form planetesimals
faster.  What is unclear is whether disks surrounding 
intermediate mass stars 
(with shorter gas disk lifetimes) retain remnant gas needed to 
damp the eccentricities of forming planetesimals 
to create planetary systems like our own 
({\it Kominami and Ida}, 2002).  
Yet the planetesimal growth time in disks surrounding 
low mass stars and brown dwarfs might be prohibitive
given the low surface densities of solids (see however 
{\it Beaulieu et al.}, 2006).  Perhaps, 
just like Goldilocks, we will find that terrestrial 
planets in stable circular orbits are found in abundance around 
sun--like stars from 0.3--3 AU.  Whether these planets have liquid water
and the potential for life as we know it to develop 
will depend on many factors (see chapter by {\it Gaidos and Selsis}). 
As results from Spitzer and other facilities 
continue to guide our understanding in the coming
years, we can look forward to steady progress.
Hopefully, new observational 
capabilities and theoretical insights will provide answers
to some of these questions at PPVI. 

{\bf Acknowledgements} We would like to the thank the referee for helpful 
comments that improved the manuscript, as well as
the conference organizers and manuscript editors
for their efforts.  MRM and DB would like to acknowledge
members of the FEPS project for their continued 
collaboration (in particular J.S. Kim and F. Fan 
for assistance with Figure 5) supported through a grant from JPL.
MRM is supported in part through the LAPLACE node 
of NASA's Astrobiology Institute. 

\centerline\textbf{REFERENCES}
\bigskip
\parskip=0pt
{\small
\baselineskip=11pt
\refs Andrews, S. and Williams, J. (2005) {\it Astrophys. J., 631}, 1134-1106.

\refs Apai D., Pascucci I., Bouwman J., Natta A., Henning T., and Dullemond C.~P.
(2005) {\it Science, 310}, 834-836.

\refs Ardila D., Golimowski, D., Krist, J., Clampin, M., Williams, J. et al. (2004) {\it Astrophys. J., 617}, L147-L150.

\refs Ardila D., Lubow, S., Golimowski, D., Krist, J., Clampin, M. et al. (2005) {\it Astrophys. J., 627}, 986-1000.

\refs Artymowicz P., Burrows C., and Paresce F. (1989) {\it Astrophys. J., 337}, 494-513.

\refs Augereau J.~C., Nelson R.~P., Lagrange A.~M., Papaloizou J.~C.~B.,
and Mouillet D. (2001) {\it Astron. Astrophys., 370}, 447-455.

\refs Aumann H., Beichman, C., Gillet, F., de Jong, T., Houck, J. et al. (1984) {\it Astrophys. J., 278}, L23-L27.

\refs Backman D. and Paresce F. (1993) In {\it Protostars and Planets III}
(E. H. Levy and J. I. Lunine, eds.), pp. 1253-1304. Univ. of Arizona, Tucson.

\refs Backman D.~E., Dasgupta A., and Stencel R.~E. (1995) {\it Astrophys. J., 450}, L35-L38.

\refs Beaulieu J.-P., Bennett, D., Fouquˆ©, P., Williams, A., Dominik, M. et al. (2006) {\it Nature,  439}, 437-440.

\refs Beckwith S.~V.~W. and Sargent A.~I. (1996) {\it Nature, 383}, 139-144.

\refs Beckwith S.~V.~W., Sargent A.~I., Chini R.~S., and Guesten R.
(1990) {\it Astron. J., 99}, 924-945.

\refs Beichman C., Bryden, G., Rieke, G., Stansberry, J., Trilling, D. et al. (2005a) {\it Astrophys. J., 622}, 1160-1170.

\refs Beichman C., Bryden, G., Gautier, T., Stapelfeldt, K., Werner, M. et al.  (2005b) {\it Astrophys. J., 626}, 1061-1069.

\refs Benz W. and Asphaug E. (1999) {\it Icarus, 142}, 5-20.

\refs Bottke W.~F., Durda D.~D., Nesvorn{\'y} D., Jedicke R., Morbidelli, A. et al. (2005) {\it Icarus, 179}, 63-94.

\refs  Brandeker A., Liseau R., Olofsson G., and Fridlund M. (2004)
{\it Astron. Astrophys., 413}, 681-691.

\refs  Bryden G., Beichman, C., Trilling, D., Rieke, G., Holmes, E. et al. (2006) {\it Astrophys. J., 636}, 1098-1113.

\refs Burns J.~A., Lamy P.~L., and Soter S. (1979) {\it Icarus, 40}, 1-48.

\refs Calvet N.,  D'Alessio, P., Watson, D., Franco-Hernandez, R., Furlan, E. et al. (2005) {\it Astrophys. J., 630}, L185-L188.

\refs Carpenter J.~M., Wolf S., Schreyer K., Launhardt R., and Henning T.
(2005) {\it Astron. J., 129}, 1049-1062.

\refs Chen C.~H. (2002), {\it B.A.A.S., 34}, 1145.

\refs  Chen C.~H., Jura M., Gordon K.~D., and Blaylock M. (2005) {\it Astrophys. J., 623}, 493-501.

\refs Cuzzi J.~N., Lissauer J.~J., Esposito L.~W., Holberg J.~B.,
Marouf E.~A., Tyler G.~L., and Boishchot A. (1984) in {\em IAU Colloq.~75: Planetary Rings}, pp. 73-199.  Univ. of Arizona, Tucson.

\refs Davis D.~R., Chapman C.~R., Weidenschilling S.~J., and Greenberg R.
(1985) {\it Icarus, 63}, 30-53.

\refs Decin G., Dominik C., Waters L.~B.~F.~M., and Waelkens C.
(2003) {\it Astrophys. J., 598}, 636-644.

\refs Dent W.~R.~F., Greaves J.~S., and Coulson I.~M.
(2005) {\it Mon. Not. R. Astron. Soc., 359}, 663-676.

\refs Dermott S.~F., Durda D.~D., Grogan K., and Kehoe T.~J.~J.
(2002) In {\em Asteroids III} (W. F. Bottke Jr. et al., eds.), pp. 423-442.
Univ. of Arizona, Tucson.

\refs Dohnanyi J.~W. (1969) {\it J. Geophys. Res., 74}, 2531-2554.

\refs Dominik C. and Decin G. (2003) {\it Astrophys. J., 598}, 626-635.

\refs Draine B. and Lee H. (1984) {\it Astrophys. J., 285}, 89-108.

\refs Duquennoy A., and Mayor M. (1991) {\it Astron. Astrophys., 248}, 485-524.

\refs Farley K.~A., Ward P., Garrison G.,
and Mukhopadhyay S. (2005) {\it Earth and Planetary Science Letters, 240}, 265-275.

\refs Fernandez, R., Brandeker, A., and Wu, Y. 
(2006) {\it Astrophys. J., 643}, 509-522. 

\refs Fischer D. and Valenti, J. (2005) {\it Astrophys. J., 622}, 1102-1117.

\refs Forrest W., Sargent, B., Furlan, E., D'Alessio, P., Calvet, N. et al. (2004) {\it Astrophys. J. Suppl., 154}, 443-447.

\refs Fujiwara A., Cerroni P., Davis D., Ryan E., and di Martino M.
(1989) In {\em Asteroids II} (R.P. Binzel et al., eds.), pp. 240-265.

\refs Golimowski D.~A.,
Ardila D.~R., Clampin M., Krist J.~E., Ford H.~C., Illingworth G.~D.
et al. (2005) In {\it Protostars and Planets V Poster Proceedings} \\
http://www.lpi.usra.edu/meetings/ppv2005/pdf/8488.pdf.

\refs Gomes R., Levison H.~F., Tsiganis K., and
Morbidelli A. (2005) {\it Nature, 435}, 466-469.

\refs Gorti U. and Hollenbach D. (2004) {\it Astrophys. J., 613}, 424-447.

\refs Greaves J. and Wyatt M. (2003)
{\it Mon. Not. R. Astron. Soc., 345}, 1212-1222.

\refs Greaves J., Holland, W., Moriarty-Schieven, G., Jenness, T.,
Dent, W. et al. (1998) {\it Astrophys. J., 506}, L133-L137.

\refs Greaves J.~S., Wyatt M.~C., Holland W.~S., and Dent W.~R.~F.
(2004) {\it Mon. Not. R. Astron. Soc., 351}, L54-L58.

\refs Greaves J., Holland, W., Wyatt, M., Dent, W., Robson, E., et al.
(2005) {\it Astrophys. J., 619}, L187-L190.

\refs Greaves J.~S., Fischer D.~A., and Wyatt M.~C. (2006)
{\it Mon. Not. R. Astron. Soc., 366}, 283-286.

\refs Grogan K., Dermott S.~F., and Durda D.~D.
(2001) {\it Icarus, 152}, 251-267.

\refs Gustafson B.  (1994) {\it Ann. Rev. Earth and Planetary Sciences, 22}, 553-595.

\refs Habing H., Dominik, C., Jourdain de Muizon, M., Kessler, M., Laureijs, R., et al. (1999) {\it Nature, 401}, 456-458.

\refs Habing H.,  Dominik, C., Jourdain de Muizon, M., Laureijs, R., Kessler, M., et al. (2001) {\it Astron. Astrophys., 365}, 545-561.

\refs Haisch K., Lada E., and Lada, C. (2001) {\it Astrophys. J., 553}, L153-L156.

\refs Hartigan P., Edwards S., and Ghandour L. (1995) {\it Astrophys. J., 452}, 736-768.

\refs Hartmann L., Calvet N., Gullbring E., and D'Alessio, P. (1998)
{\it Astrophys. J., 495}, 385-400.

\refs Harvey P.~M. (1985) In {\em Protostars and Planets II}
(B. Matthews et al., eds.), pp. 484-492. Univ. of Arizona, Tucson.

\refs Hillenbrand, L., Strom, S., Calvet, N., Merrill, K., Gatley, I.
et al. (1998) {\it Astron. J., 116}, 1816-1841.

\refs Hines D., Backman, D., Bouwman, J., Hillenbrand, L., Carpenter, J. et al. (2006) {\it Astrophys. J., 638}, 1070-1079.

\refs Holland W., Greaves, J., Zuckerman, B., Webb, R., McCarthy, C. et al. (1998) {\it Nature, 392}, 788-790.

\refs Holland W., Greaves, J., Dent, W., Wyatt, M., Zuckerman, B. et al. (2003) {\it Astrophys. J., 582}, 1141-1146.

\refs Hollenbach D., Gorti, U., Meyer, M., Kim, J., Morris, P. et al. (2005) {\it Astrophys. J., 631}, 1180-1190.

\refs Jayawardhana R., Fisher S., Hartmann L., Telesco C., Pina R.,
and Fazio G. (1998) {\it Astrophys. J., 503}, L79-L82.

\refs Jayawardhana R., Holland W., Kalas P., Greaves J., Dent W. et al.,  (2002) {\it Astrophys. J., 570}, L93-L96.

\refs Jensen E.~L.~N., Mathieu R.~D., and Fuller G.~A. (1996)
{\it Astrophys. J., 458}, 312-326.

\refs Jewitt D.~C. (1994) {\it Astron. J., 108}, 661-665.

\refs Jonkheid B., Kamp I., Augereau J.-C., and
van Dishoeck, E.~F. (2005) In {\em Astrochemistry Throughout the Universe}
(D. Lis et al., eds.), pp. 49. Cambridge University Press, Cambridge.

\refs Jura M., Malkan M., White R., Telesco C., Pina R.,
and Fisher R.~S. (1998) {\it Astrophys. J. 505}, 897-902.

\refs Jura M., Chen, C., Furlan, E., Green, J., Sargent, B. et al. (2004) {\it Astrophys. J. Suppl., 154}, 453-457.

\refs Kalas P. (2005) {\it Astrophys. J., 635}, L169-L172.

\refs Kalas P. and Jewitt D. (1995) {\it Astron. J., 110}, 794-804.

\refs Kalas P., Graham J.., Beckwith S., Jewitt D.,
and Lloyd J. (2002) {\it Astrophys. J., 567}, 999-1012.

\refs Kalas P., Liu M., and
Matthews B. (2004) {\it Science, 303}, 1990-1992.

\refs Kalas P., Graham J., and Clampin, M. (2005)
{\it Nature, 435}, 1067-1070.

\refs  Kalas P., Graham J., Clampin M.~C., and Fitzgerald M.~P.
(2006) {\it Astrophys. J., 637}, L57-L60.

\refs Kenyon S. and Bromley B. (2002a) {\it Astron. J., 123}, 1757-1775.

\refs Kenyon S. and Bromley B. (2002b) {\it Astrophys. J., 577}, L35-L38.

\refs Kenyon S. and Bromley B. (2004) {\it Astron. J., 127}, 513-530.

\refs Kenyon S. and Bromley B. (2005) {\it Astron. J., 130}, 269-279.

\refs Kenyon S., and Hartmann, L. (1995) {\it Astrophys. J. Suppl., 101},
117-171.

\refs Kim J., Hines, D., Backman, D., Hillenbrand, L., Meyer, M. et al. (2005) {\it Astrophys. J., 632}, 659-669.

\refs Klahr H. and Lin D. (2001) {\it Astrophys. J., 554}, 1095-1109.

\refs Koerner D., Ressler M., Werner M., and
Backman D. (1998) {\it Astrophys. J., 503}, L83-L87.

\refs Koerner D., Jensen E., Cruz K., Guild T.,
and Gultekin K. (2000) {\it Astrophys. J., 533}, L37-L40.

\refs Koerner D., Sargent A., and Ostroff N. (2001) {\it Astrophys. J., 560}, L181-L184.

\refs Kominami J., and Ida S. (2002) {\it Icarus, 157}, 43-56.

\refs Krist J., Ardila, D., Golimowski, D., Clampin, M., Ford, H. 
et al. (2005) {\it Astron. J., 129}, 1008-1017.

\refs Krivov A., Mann I., and Krivova N. (2000)
{\it Astron. Astrophys., 362}, 1127-1137.

\refs Lagrange A.-M., Beust, H., Mouillet, D., Deleuil, M., 
Feldman, P. et al. (1998) {\it Astron. Astrophys., 330}, 1091-1108.

\refs Lagrange A., Backman D. and
Artymowicz, P. (2000) In {\it Protostars and Planets IV} 
(V. Mannings et al., eds.), pp. 639-672. Univ. of Arizona, Tucson. 

\refs Landgraf M., Liou J.-C., Zook H., and
Gr{\"u}n E. (2002) {\it Astron. J., 123}, 2857-2861.

\refs Larwood J., and Kalas P.
(2001) {\it Mon. Not. R. Astron. Soc., 323}, 402-416.

\refs Lawson W., Lyo, A., and Muzerolle, J. 
(2004) {Mon. Not. R. Astron. Soc., 351}, L39-L43. 

\refs Lecavelier des Etangs A., Vidal-Madjar, A., 
Roberge, A., Feldman, P., Deleuil, M.
 et al. (2001) {\it Nature, 412}, 706-708.

\refs Levison H.~F. and Morbidelli, A.
(2003) {\it Nature, 426}, 419-421.

\refs Liou J.-C. and Zook H. (1999) {\it Astron. J., 118}, 580-590.

\refs Lissauer J., and Griffith C. (1989) {\it Astrophys. J., 340}, 468-471.

\refs Liu M.~C. (2004) {\it Science, 305}, 1442-1444.

\refs Liu M.~C., Matthews B.~C., Williams J.~P., and Kalas P.~G. (2004)
{\it Astrophys. J., 608}, 526-532.

\refs Love S.~G. and Brownlee, D.~E. (1993) {\it Science, 262}, 550-552.

\refs Low F., Hines D., and Schneider G. (1999)
{\it Astrophys. J., 520}, L45-L48.

\refs Low F., Smith P., Werner M., Chen C., Krause V. et al.
(2005) {\it Astrophys. J., 631}, 1170-1179.

\refs Malhotra R. (1993) {\it Nature, 365}, 819-821.

\refs Mamajek E., Meyer M., Hinz P., Hoffmann W., Cohen M.,
and Hora, J. (2004) {\it Astrophys. J., 612}, 496-510.

\refs Marcy G., Cochran W., and Mayor, M. (2000)
In {\em  Protostars and Planets IV} 
(V. Mannings et al., eds.), pp. 1285-1311.
Univ. of Arizona, Tucson. 

\refs Marengo, M. et al.
(2005) In {\it Protostars and Planets V Poster Proceedings} \\
http://www.lpi.usra.edu/meetings/ppv2005/pdf/8566.pdf.

\refs Marsh K., Velusamy T., Dowell C., Grogan K., and
Beichman, C. (2005) {\it Astrophys. J., 620}, L47-L50.

\refs Metchev S., Eisner J., Hillenbrand L., and
Wolf, S. (2005) {\it Astrophys. J., 622}, 451-462.

\refs Meyer M.~R., and Beckwith S. (2000)
In {\it LNP Vol.~548: ISO Survey of a Dusty Universe} 
(D. Lemke et al., eds.), pp, 341-352. 
Springer-Verlag, Heidelberg. 

\refs Meyer M.,  Hillenbrand, L., Backman, D., Beckwith, S., 
 Bouwman, J. et al. (2004) {\it Astrophys. J. Suppl., 154}, 422-427.

\refs  Moro-Mart{\'{\i}}n A. and Malhotra, R. (2002) {\it Astron. J., 124},
2305-2321.

\refs Muzerolle J., Hillenbrand L., Calvet N., Brice{\~n}o C.,
and Hartmann L. (2003) {\it Astrophys. J., 592}, 266-281.

\refs Nagasawa M., Lin D., and Thommes E. (2005)
{\it Astrophys. J., 635}, 578-598.

\refs Najita J. and Williams J. (2005) {\it Astrophys. J., 635}, 625-635.

\refs Natta A., Grinin V., and Mannings V. (2000), In
{\it Protostars and Planets IV} (V. Mannings et al., eds.), 
pp. 559-587. Univ. of Arizona, Tucson. 

\refs Nesvorn{\'y} D., Bottke W., Levison H., and Dones L.
(2003) {\it Astrophys. J., 591}, 486-497.

\refs Okamoto Y.,  Kataza, H., Honda, M., Yamashita, T., Onaka, T. et al. (2004) {\it Nature, 431}, 660-663.

\refs Plavchan P., Jura M., and Lipscy S. (2005) {\it Astrophys. J., 631}, 1161-1169.

\refs Raymond S., Quinn T., and Lunine J. (2004) {\it Icarus, 168}, 1-17.

\refs Rieke G., Su, K., Stansberry, J., Trilling, D., 
Bryden, G. et al. (2005) {\it Astrophys. J., 620}, 1010-1026.

\refs Richter, M., Jaffe, D., Blake, G., \& Lacy, J. (2002)
{\it Astrophys. J., 572}, L161-L164.

\refs Roberge A., Weinberger A., Redfield S., and Feldman P. (2005)
{\it Astrophys. J., 626}, L105-L108.

\refs Roberge, A., Feldman, P., Weinberger, A., Deleuil, M., Bouret, J.
(2006) {\it Nature, 441} 724-726. 

\refs Roques F., Scholl H., Sicardy B., and Smith B. (1994) {\it Icarus, 108}, 37-58.

\refs Rydgren A. (1978) In {\em Protostars and Planets} 
(T. Gehrels, ed.), pp, 690-698. Univ. of Arizona, Tucson. 

\refs Schneider G., Smith, B., Becklin, E., Koerner, D., Meier, R. et al. (1999) {\it Astrophys. J., 513}, L127-L130.

\refs Schneider G., Becklin E., Smith B., Weinberger A., Silverstone M.,
and Hines D. (2001) {\it Astron. J., 121}, 525-537.

\refs Schneider G., Silverstone M.~D., and Hines D.~C. (2005)
{\it Astrophys. J., 629}, L117-L120.

\refs Sheret I., Ramsay Howat S., and Dent W.
(2003) {\it Mon. Not. R. Astron. Soc., 343}, L65-L68.

\refs Shu, F., Tremaine, S., Adams, F., and Ruden, S.
(1990) {\it Astrophys. J., 358}, 495-514.  

\refs Silverstone M., Meyer, M., Mamajek, E., Hines, D., Hillenbrand, L. et al. (2006) {\it Astrophys. J., 639}, 1138-1146.

\refs Simon M. and Prato L. (1995) {\it Astrophys. J., 450}, 824-829.

\refs Skrutskie M., Dutkevitch D., Strom S., Edwards S., Strom K.,
and Shure M. (1990) {\it Astron. J., 99}, 1187-1195.

\refs Smith B. and Terrile R. (1984) {\it Science, 226}, 1421-1424.

\refs Song I., Zuckerman B., Weinberger A., and
Becklin, E. (2005) {\it Nature, 436}, 363-365.

\refs Spangler C., Sargent A., Silverstone M., Becklin E.,
and Zuckerman, B. (2001) {\it Astrophys. J., 555}, 932-944.

\refs Stapelfeldt K., Holmes, E., Chen, C., Rieke, G., Su, K., et al. (2004) {\it Astrophys. J. Suppl., 154}, 458-462.

\refs Stauffer J. (2004)
In {\em ASP Conf.~Ser.~324: Debris Disks and the Formation of Planets, 324} 
(L. Caroff et al., eds.), pp. 100-111.  Astronomical Society of the Pacific, 
San Francisco. 

\refs Stern S. (1996) {\it Astron. J., 112}, 1203-1211.

\refs Stern S., and Colwell J. (1997) {\it Astrophys. J., 490}, 879-882.

\refs Strom R., Malhotra R., Ito T., Yoshida F.,
and Kring, D. (2005) {\it Science, 309}, 1847-1850.

\refs Strom S., Edwards S., and Skrutskie M.
(1993) In {\em Protostars and Planets III} 
(E. Levy and J. Lunine, eds.), pp. 837-866.
Univ. of Arizona, tucson. 

\refs Su K., Rieke, G., Misselt, K., Stansberry, J., Moro-Martin, A. et al. 
(2005) {\it Astrophys. J., 628}, 487-500.

\refs Takeuchi T. and Artymowicz P. (2001) {\it Astrophys. J., 557}, 990-1006.

\refs Takeuchi T., Clarke C., and Lin D. (2005) {\it Astrophys. J., 627}, 286-292.

\refs Tanaka H., Inaba S., and Nakazawa, K. (1996) {\it Icarus, 123}, 450-455.

\refs Telesco C., Fisher, R., Piña, R., Knacke, R., Dermott, S. 
 et al. (2000) {\it Astrophys. J., 530}, 329-341.

\refs Telesco C., Fisher, R., Wyatt, M., Dermott, S., 
Kehoe, T. et al. (2005) {\it Nature, 433}, 133-136.

\refs Teplitz V., Stern S., Anderson J., Rosenbaum D., Scalise R.,
and Wentzler, P. (1999) {\it Astrophys. J., 516}, 425-435.

\refs Thi W., Blake, G., van Dishoeck, E., van Zadelhoff, G., 
Horn, J. et al. (2001a) {\it Nature, 409}, 60-63.

\refs Thi W., van Dishoeck, E., Blake, G., van Zadelhoff, G., Horn, J.
et al. (2001b) {\it Astrophys. J., 561}, 1074-1094.

\refs Th{\'e}bault P., Augereau J., and Beust, H. (2003) {\it Astron. Astrophys., 408}, 775-788.

\refs Th{\'e}bault P. and Augereau J.-C. (2005) {\it Astron. Astrophys., 437}, 141-148.

\refs Thommes E., Duncan M., and Levison H.
(1999) {\it Nature, 402}, 635-638.

\refs  Weidenschilling S., Spaute D., Davis D., Marzari F.,
and Ohtsuki K. (1997) {\it Icarus, 128}, 429-455.

\refs Weinberger A., Becklin E., and Zuckerman, B. (2003)
{\it Astrophys. J., 584}, L33-L37.

\refs Weinberger A., Becklin E., Zuckerman B., and Song, I.
(2004) {\it Astron. J., 127}, 2246-2251.

\refs Whitmire D., Matese J., and Whitman P. (1992)
{\it Astrophys. J., 388}, 190-195.

\refs Williams J., Najita J., Liu M., Bottinelli S., Carpenter J. et al. (2004) {\it Astrophys. J., 604}, 414-419.

\refs Wilner D., Holman M., Kuchner M., and Ho P.
(2002) {\it Astrophys. J., 569}, L115-L119.

\refs Wisdom J. (1980) {\it Astron. J., 85}, 1122-1133.

\refs Wolk S. and  Walter F. (1996) {\it Astron. J., 111}, 2066-2076.

\refs Wyatt M. (2003) {\it Astrophys. J., 598}, 1321-1340.

\refs Wyatt M. (2005) {\it Astron. Astrophys., 433}, 1007-1012.

\refs Wyatt M. (2006) {\it Astrophys. J., 639}, 1153-1165. 

\refs Wyatt M. and Dent W. (2002) {\it Mon. Not. R. Astron. Soc., 334}, 589-607.

\refs Wyatt M., Dermott S., Telesco C., Fisher R., Grogan K. et al. (1999) {\it Astrophys. J., 527}, 918-944.

\refs Wyatt M., Dent W., and Greaves J. (2003)
{\it Mon. Not. R. Astron. Soc., 342}, 876-888.

\refs Wyatt M., Greaves J., Dent W., and Coulson I. (2005)
{\it Astrophys. J., 620}, 492-500.

\refs Zuckerman B. (2001) {\it Ann. Rev. Astron. Astrophys., 39}, 549-580.

\refs Zuckerman B. and Becklin E. (1993) {\it Astrophys. J. 414}, 793-802.

\refs Zuckerman B., Forveille T., and Kastner J.. (1995) {\it Nature, 373}, 494-496.
}
\end{document}